\documentclass{LMCS}

\usepackage{enumerate}
\usepackage{hyperref}

\theoremstyle{plain}

\def\eqalign#1{\null\,\vcenter{\openup\jot\mathsurround=0 pt
  \ialign{\strut\hfil$\displaystyle{##}$&$\displaystyle{{}##}$\hfil
      \crcr#1\crcr}}\,}
\def\ord{\mathop{\rm ord}\nolimits}
\def\iff {if\null f\ }
\def\<{\langle}                         \def\>{\rangle}

\def\doi{5 (2:6) 2009}
\lmcsheading%
{\doi}
{1--21}
{}
{}
{Feb.~21, 2008}
{Apr.~27, 2009}
{}   

\begin{document}

\title[The Omega Rule is $\mathbf{\Pi_{1}^{1}}$-Complete]{The Omega Rule is $\mathbf{\Pi_{1}^{1}}$-Complete in the $\lambda\beta$-Calculus}

\author[B.~Intrigila]{Benedetto Intrigila\rsuper a}   
\address{{\lsuper a}Universit\`{a} di Roma "Tor Vergata" \\
          Rome, Italy} 
\email{intrigil@mat.uniroma2.it}  

\author[R.~Statman]{Richard Statman\rsuper b} 
\address{{\lsuper b}Carnegie-Mellon University \\
Pittsburgh, PA, USA}    
\email{rs31@andrew.cmu.edu}  




\begin{abstract}
  \noindent In a functional calculus, the so called
$\omega$-rule states that if two terms $P$ and $Q$ applied to any
closed term $N$ return the same value (i.e. $PN = QN$), then they
are equal (i.e. $P = Q$ holds). As it is well known, in the
$\lambda\beta$-calculus the $\omega$-rule does not hold, even when
the $\eta$-rule (weak extensionality) is added to the calculus. A
long-standing problem of H. Barendregt (1975) concerns the
determination of the logical power of the $\omega$-rule when added
to the $\lambda\beta$-calculus. In this paper we solve the problem,
by showing that the resulting theory is
$\mathbf{\Pi_{1}^{1}}$-Complete.
\end{abstract}

\maketitle

\section*{Introduction}\label{S:one}

In a functional calculus, the so called $\omega$-rule states that if
two terms $P$ and $Q$ applied to any closed term $N$ return the same
value (i.e. $PN = QN$), then they are equal (i.e. $P = Q$ holds). As
it is well known, in the $\lambda\beta$-calculus the $\omega$-rule
does not hold, even when the $\eta$-rule (weak extensionality) is
added to the calculus.

It is therefore natural to investigate the logical status of
the $\omega$-rule in $\lambda$-theories.

We have first considered constructive forms of such rule in
\cite{IS2002}, obtaining r.e. $\lambda$-theories which are closed
under the $\omega$-rule. This gives the counterintuitive result that
closure under the $\omega$-rule does not necessarily give rise to
non constructive $\lambda$-theories,
thus solving a problem of A. Cantini (see \cite{Cantini}).

Then we have considered the $\omega$-rule with respect to the highly
non constructive $\lambda$-theory $\mathcal{H}$. The theory
$\mathcal{H}$ is obtained  extending $\beta$-conversion by
identifying all closed unsolvables. $\mathcal{H}\mathbf{\omega}$ is
the closure of this theory under the $\omega$-rule (and
$\beta$-conversion). A long-standing conjecture of H. Barendregt
(\cite{Ba84}, Conjecture 17.4.15) stated that the provable equations
of $\mathcal{H}\mathbf{\omega}$ form a
$\mathbf{\Pi_{1}^{1}}$-Complete set. In \cite{IS2006},
we solved in the affirmative the problem.

Of course the most important problem is to determine the logical
power of $\omega$-rule when added to the {\em pure}
$\lambda\beta$-calculus.

As in \cite{Ba84}, we call $\lambda\omega$ the theory that results
from adding the $\omega$-rule to the {\em pure}
$\lambda\beta$-calculus. In \cite{IS2004}, we showed that the
$\lambda\omega$ is {\em not} recursively enumerable, by giving a
many-one reduction of the set of true $\mathbf{\Pi_{2}^{0}}$ sentences
to the set of closed equalities provable in $\lambda\omega$, thus
solving a problem originated with H.  Barendregt and re-raised in
\cite{FlaggMyhill}.

The problem of the logical upper bound to $\lambda\omega$ remained
open. That this bound is $\mathbf{\Pi_{1}^{1}}$ has been conjectured
again by H. Barendregt in the well known Open Problems List, which
ends the 1975 Conference on "$\lambda$-Calculus and Computer Science
Theory", edited by C. B\"{o}hm \cite{Boh}. Here we solve in the
affirmative this conjecture. The celebrated {\em Plotkin terms}
(introduced in \cite{Plo}) furnish the main technical tool.

\subsection{Remarks on the Structure of the Proof. }
The present paper is a revised and improved version of
\cite{IS2007}. It is self-contained, with the exception of some
specific points where we use results and methods from \cite{IS2004}.
Such points will be precisely indicated in Section \ref{PT} and in
Section \ref{BarConstr}. The authors are working to a comprehensive
formalism to give a unified presentation of all the results.  At
present, however, this could not have been done without great
complications.

To help the reader, we now describe in an informal way the general
idea of the proof.

As already for the result in \cite{IS2004}, the proof relies on
suitable modifications of the mentioned Plotkin terms. Roughly
speaking, Plotkin's construction gives rise, in the usual
$\lambda\beta\eta$-calculus, to pairs of closed terms $P_0$ and $P_1$
such that for every closed term $M$, $P_0 M$ and $P_1 M$ are
$\beta\eta$-convertible. On the other hand, $P_0$ and $P_1$ are {\em
  not} themselves $\beta\eta$-convertible (see \cite{Ba84}, 17.3.26).

When we add the $\omega$-rule to the $\lambda\beta$-calculus, such
terms - suitably modified - become a way to express various forms of
{\em universal quantification}. Intuitively, $P_0$ and $P_1$ are equal
if and only if {\em for all} $M$ belonging to some given set of terms,
$P_0 M$ and $P_1 M$ are equal.

There are two points that must be stressed.
\begin{enumerate}[$\bullet$]\item First,
different quantifiers require different specific constructions of
suitable Plotkin terms.
\item Second, to properly use equality between $P_0$ and $P_1$ as a test for quantification, one must
 exclude that $P_0 M = P_1 M$ holds for some $M$ not belonging to
 the set of interest.
\end{enumerate}
Focusing on the second problem, the technical tool that we have used
- both in \cite{IS2004} and in the present paper - is to cast proofs
in the $\lambda\beta$-calculus with the $\omega$-rule, in some kind
of "normal form". (Observe that, in presence of the $\omega$-rule,
proofs
become infinitary objects.)
In particular as "normal form" for proofs, we have used in
\cite{IS2004} the notion of {\em cascaded proof}. Here we use the
notion of {\em canonical proof} introduced in Section
\ref{canonical7}.
In both cases, the intuitive idea is to extensively use the
$\omega$-rule to limit the use of $\beta$-reductions. This makes the
behavior of the (various) Plotkin terms more controllable, which, in
turn, makes the
mentioned problem solvable.
It turns out that one cannot use a unique "normal form", or at least
we were not able to do this. In particular, observe that we need
terms for two kinds of quantifier:

\begin{enumerate}[$\bullet$]\item Arithmetical quantification over 
recursive enumerable sets of terms.
\item One second order universal quantification
 to express $\mathbf{\Pi_{1}^{1}}$-complete problems.
\end{enumerate}

Different kinds of terms are used to express, {\em via} their
equality, the two kinds of quantification. So in the present paper
we concentrate on {\em canonical proofs}. This kind of "normal form"
is suitable to cope with terms whose equality is used to express a
$\mathbf{\Pi_{1}^{1}}$-complete problem. It is not suitable,
however, to properly control the behavior of terms related to
first-order quantification. For such terms, we rely on the methods
used in \cite{IS2004} for the analysis of {\em cascaded proofs}.
The general scheme of the proof is as follows:
\begin{enumerate}[$\bullet$]
\item
In Section \ref{canonical7}, we introduce the notion of {\em
canonical proof} and prove that every provable equality has a
canonical proof.
\item  In Section \ref{PT}, we introduce suitable Plotkin terms to
express quantification over Church numerals.
\item  In Section \ref{BarConstr}, we introduce suitable Plotkin terms to
express second order quantification on sequences of numbers to
reduce the $\mathbf{\Pi_{1}^{1}}$-complete problem of
well-foundedness of recursive trees to equality of terms in the
$\lambda\beta$-calculus with the $\omega$-rule.
\end{enumerate}

\section{The $\omega$-rule}\label{two}

Notation will be standard and we refer to \cite{Ba84}, for
terminology and results on $\lambda$-calculus. In particular:
\begin{enumerate}[$\bullet$]
\item $\equiv$ denotes syntactical identity;
\item $\longrightarrow_{\beta}$, $\longrightarrow_{\eta}$ and $\longrightarrow_{\beta\eta}$
 denote  $\beta$-, $\eta$- and, respectively, $\beta\eta$-reduction and
$\longrightarrow^{*}_{\beta}$, $\longrightarrow^{*}_{\eta}$ and
$\longrightarrow^{*}_{\beta\eta}$ their respective reflexive and
transitive closures;
\item $=_{\beta}$ and $=_{\beta\eta}$ denote  $\beta$- and, respectively, ${\beta\eta}$-conversion;
\item combinators (i.e. closed $\lambda$-terms)  such e.g. $\mathbf{I}$ have the usual meaning;
\item $\underline{k}$ denotes the kth Church numeral.\end{enumerate}
$\lambda$-terms are denoted by capital letters: in particular we
adopt the convention that $F$, $G$, $H$, $J$, $M$, $N$, $P$,
$Q,\ldots$ are {\em closed} terms and $U,V,X,Y,W,Z$ are possibly
 {\em open} terms.

For a $\lambda$-term the notion of {\em having order $0$} has the
usual meaning (\cite{Ba84} 17.3.2). We shall also call {\em
zero-term} a term of order $0$. As usual, we say that a term has
{\em positive order} if it is not of order zero.
 We shall refer to a $\beta$-reduction performed not within the
scope of a $\lambda$ as {\em a weak $\beta$-reduction}. In the
sequel, we shall need the following notions. We define the notions
of {\em trace} and {\em extended trace} (shortly {\em etrace}) as
follows. Given the reduction $F \longrightarrow_{\beta\eta}^{*} G$
and the closed subterm $M$ of $F$ the {\em traces} of $M$ in the
terms of the reduction are simply the copies of $M$ until each is
either
 deleted by a contraction of a redex with a dummy $\lambda$
  or altered by a reduction internal to $M$ or by a reduction
with $M$ at the head (when $M$ begins with $\lambda$). The notion of
{\em etrace} is the same except that we allow internal reductions,
so that a copy of $M$ altered by an internal reduction continues to
be an etrace.

 By $\lambda \beta$ we denote the theory of
$\beta$-convertibility (see \cite{Ba84}). The theory $\lambda
\omega$ is obtained by adding the so called {\em $\omega$-rule} to
$\lambda \beta$, see \cite{Ba84} 4.1.10.

We formulate $\lambda\omega$ slightly differently. In particular, we
want a formulation of the theory such that only equalities between
{\em closed} terms can be proven. Moreover it will be convenient to
use $\beta\eta$-conversion. The so called $\eta$-rule (that is
$(\lambda x.Mx) =_{\eta} M$)
 obviously holds in $\lambda\omega$. Nevertheless it will be useful to have
 this rule at disposal to put proofs in some specified forms.

\begin{defi}
{\em Equality in  $\lambda\omega$ } (denoted by $=_{\omega}$) is
defined by the following rules:
\begin{enumerate}[$\bullet$]
\item $\beta\eta$-conversion:
\begin{center}
                       if $M =_{\beta\eta} N$ then $M =_{\omega} N$
\end{center}\smallskip

\item the rule of substituting equals for equals in the form:
\begin{center}
    if $M =_{\omega} N$ then $PM =_{\omega} PN$
\end{center}\smallskip

\item transitivity and symmetry of equality,
\item the $\omega$-rule itself:
  \[ \frac{ \forall M,\ M \ closed,\  PM =_{\omega} QM}{P~=_{\omega}~Q}\]
\end{enumerate}
\end{defi}

We leave to the reader to check that the formulation above is
equivalent to the standard one (see Chapter 4 of \cite{Ba84}).

As usual proofs in $\lambda\omega$ can be thought of as (possibly
infinite) well-founded trees. In particular the tree of a proof
either ends with an instance of the $\omega$-rule or has an end
piece consisting of a finite tree of equality inferences all of
whose leaves are either $\beta\eta$ conversions or direct
conclusions of the $\omega$-rule. It is easy to see that each such
endpiece can be put in the form:
      $$F =_{\beta\eta} G_1 M_1 =_{\omega} G_1 N_1 =_{\beta\eta}
      G_2 M_2 =_{\omega} G_2 N_2 =_{\beta\eta}
      \dots G_t M_t =_{\omega} G_t N_t =_{\beta\eta} H$$
where $M_i =_{\omega}N_i$, for $1 \leq i \leq t$ are direct
conclusions of the $\omega$-rule. See \cite{IS2006}, Section 5, for
more details. While the context is slightly different, the argument
is {\em verbatim} the same. This is a particular case of a general
result due to the second author of the present paper, see
\cite{St1}. Moreover, by the Church-Rosser Theorem this
configuration of inferences can be put in the form
\begin{equation}\label{star}
F \longrightarrow^{*}_{\beta\eta} J_1 \
_{\beta\eta}^{*}\!\longleftarrow G_1 M_1 =_{\omega} G_1 N_1
\longrightarrow^{*}_{\beta\eta}
 J_2 \ _{\beta\eta}^{*}\!\longleftarrow\end{equation}
$$\ _{\beta\eta}^{*}\!\longleftarrow G_2 M_2 =_{\omega} G_2 N_2 =_{\beta\eta}
\longrightarrow^{*}_{\beta\eta} \dots $$
$$\dots  \ _{\beta\eta}^{*}\!\longleftarrow G_t M_t =_{\omega} G_t N_t
\longrightarrow^{*}_{\beta\eta} J_{t+1} \
_{\beta\eta}^{*}\!\longleftarrow H$$
 where $M_i =_{\omega}N_i$, for $1 \leq i \leq t$, are as
above. We shall call the sequence (\ref{star}) {\em the standard
form} for the endpiece of a proof.

 Since proofs are infinite
trees (denoted by symbols $\mathcal{T}$, $\mathcal{T}'$ etc.), they
can be assigned countable ordinals. We shall
 need a few facts about countable ordinals,
that we briefly mention in the following. For the basic notions on
countable ordinals, see e.g. \cite{Sch}.\bigskip

\noindent{\bf (a) Cantor Normal Form to the Base Omega} $(\omega)$\\
Every countable ordinal $\alpha$ can be written uniquely in the form
                    $\omega^{\alpha_1}*n_1 + \dots  + \omega^{\alpha_k}*n_k$
where $n_1 , \dots  , n_k$ are positive integers and $\alpha_1 > \dots  >
\alpha_k$ are ordinals.\medskip

\noindent{\bf (b) Hessenberg Sum}\\
 Write $\alpha =  \omega^{\alpha_1}*n_1 + \dots 
+ \omega^{\alpha_k}*n_k$ and
       $\gamma =  \omega^{\alpha_1}*m_1 + \dots  + \omega^{\alpha_k}*m_k$
where some of the $n_i$ and $m_j$ may be 0. Then the {\em Hessenberg
Sum} is defined as follows: $ \alpha \oplus \gamma =_{def}
           \omega^{\alpha_1}*(n_1+m_1) + \dots  + \omega^{\alpha_k}*(n_k + m_k
           )$.

Hessenberg sum is strictly increasing on both arguments. That is,
for $\alpha, \gamma$ different
from $0$, we have: $\alpha, \gamma < \alpha \oplus \gamma$.\medskip

\noindent{\bf (c) Hessenberg Product}\\
            We only need this for product with an integer. We put:
            $\alpha \odot n =_{def} \alpha \oplus \dots  \oplus \  \alpha$
            $n$-times.\bigskip

  Coming back to proofs, observe first that we can assume that if a
  proof has an endpiece, then this endpiece is in standard form (see
  above).  The ordinal that we want to assign to a proof $\mathcal{T}$
  (considered as a tree) is the transfinite ordinal
  $\ord(\mathcal{T})$, {\em the order of } $\mathcal{T}$, defined
  recursively by
\vfill\eject

\begin{defi}\label{ord}\hfill
\begin{enumerate}[$\bullet$]
\item If
$\ \mathcal{T}$ ends in an endpiece computation of the form
(\ref{star}) with no instances of the $\omega$-rule ($t = 0$), that
is consisting of a unique $\beta\eta$-conversion, then
$\ord(\mathcal{T}) =_{def} 1$;
\item If $\ \mathcal{T}$ ends in an instance of
the $\omega$-rule whose premises have trees resp. $\mathcal{T}_{1},
\ldots \mathcal{T}_{i}, \ldots$ then $\ord(\mathcal{T}) =_{def}
\omega ^{\theta}$, with $\theta = Sup\{\ord(\mathcal{T}_1) \oplus \dots 
\oplus \ord(\mathcal{T}_i) : i=1,2,\dots \}$;
\item If
$\ \mathcal{T}$ ends in an endpiece computation of the form
(\ref{star}), with $t > 0$ instances of the $\omega$-rule,  and the
$t$ premises
 $M_1 =_{\omega}
N_1, \dots , M_t =_{\omega}N_t$
 have resp. trees $\mathcal{T}_{1}, \ldots, \mathcal{T}_{t}$
then $\ord(\mathcal{T}) =_{def} 1 \oplus \ord(\mathcal{T}_1) \oplus
\ord(\mathcal{T}_{2}) \dots  \oplus \ord(\mathcal{T}_t) $.
\end{enumerate}
 Here $\oplus$ is the Hessenberg  sum of
 ordinals defined above.
\end{defi}

We shall need also the following notion.
\begin{defi}
If $\mathcal{T}$ ends in an endpiece computation of the form
(\ref{star}), with $t > 0$ instances of the $\omega$-rule,  and the
$t$ premises
 $M_1 =_{\omega}
N_1, \dots , M_t =_{\omega}N_t$
 have resp. trees $\mathcal{T}_{1}, \ldots, \mathcal{T}_{t}$
then $rank(\mathcal{T})$, the {\em rank} of $\mathcal{T}$, is the
maximum of  $ \ord(\mathcal{T}_1)$, $\ord(\mathcal{T}_{2})$, \dots  ,
$\ord(\mathcal{T}_t) $.
\end{defi}

We need the following propositions.

\begin{prop}

If $\ \mathcal{T}$ ends in an endpiece computation of the form
(\ref{star}), with $t > 0$, and the equations $M_1 =_{\omega}
N_1$,\dots , $M_t =_{\omega}N_t$,
 have resp. trees $\mathcal{T}_{1}, \ldots, \mathcal{T}_{t}$
 then $\ord(\mathcal{T})
> \ord(\mathcal{T}_i)$, for each $i = 1, \dots , t $.
\end{prop}
\proof $\ord(\mathcal{T}_i) > 0$ and $\oplus$ is strictly increasing
on its arguments. \qed

\begin{prop}\label{Fact}
Assume that $\mathcal{T}$ ends in an instance of the $\omega$-rule
whose premises have, respectively, trees $\mathcal{T}_1, \dots ,
\mathcal{T}_t, \dots$ Then for any integers $t, n_1 , \dots ,n_t, $
\[\ord(\mathcal{T}) > \ord(\mathcal{T}_1)\odot n_1  \oplus \dots 
 \oplus  \ord(\mathcal{T}_t)\odot n_t\ .
\]
\end{prop}
\proof Let $\ord(\mathcal{T}_i) = \alpha_i$, for $1 \leq i \leq t$
and put all $\alpha_1, \dots  , \alpha_t$
into Cantor normal form:
\[\alpha_1 = \omega^{\beta_ 1} * n_{11} + \dots + \omega^{\beta_ k} * n_{1k}
\quad\dots\quad
\alpha_t = \omega^{\beta_ 1} * n_{t1} + \dots + \omega^{\beta_ k} *
n_{tk}\ .
\]
Let $n = \max \{ n_r, n_{ij} \} + 1$, with $j,r = 1 \dots  t$ and $i = 1
\dots  k$ . Then
\[\eqalign{
  \ord(\alpha_1)\odot n_1  \oplus \dots  \oplus  \ord(\alpha_t)\odot n_t
&< \ord(\alpha_1)\odot n  \oplus \dots  \oplus  \ord(\alpha_t)\odot n\cr
&= (\alpha_1 \oplus \dots  \oplus \alpha_t) \odot n\cr
&\leq \omega^{\beta_1} *n*k*t*n\ .\cr
  }
\]
Now let $\theta = Sup\{\ord(\mathcal{T}_1) \oplus \dots  \oplus
\ord(\mathcal{T}_i) : i=1,2,\dots \}$.
 We have $\omega^{\beta_1} <
\theta \leq \omega^{\theta} = \ord(\mathcal{T})$.
But $\ord(\mathcal{T})$ is a countable ordinal of the form
$\omega^\gamma$ and is thus closed under addition. Hence
$\omega^{\beta_ 1} *n*k*t*n < \ord(\mathcal{T})$.
This ends the proof.
\qed

\begin{rem} Since for proofs $\mathcal{T}$  we shall mainly use
$\ord(\mathcal{T})$, we sometimes refer to $\ord(\mathcal{T})$ simply
as {\em the ordinal} of the proof $\mathcal{T}$.
 \end{rem}
\section{Canonical Proofs}\label{canonical7}

We want to show that proofs in $\lambda\omega$ can be set in a
suitable form.

\begin{defi}
We say that $M$ has {\em the same form as} $N$ \iff
\begin{enumerate}[$\bullet$]
\item in case of $N \equiv \lambda y_1 \dots y_n. Y L_1 \dots L_m$,
  where $Y$ begins with $\lambda$, we have 
\begin{enumerate}[$-$]
\item $M \equiv \lambda y_1 \dots
  y_n. Z P_1 \dots P_m$, where $Z$ begins with $\lambda$,
\item $\lambda y_1 \dots y_n.Y =_{\omega} \lambda y_1 \dots y_n.Z$ ,
\item
  and for every $i$ with $1 \leq i \leq m$,
\[\lambda y_1 \dots y_n. L_i =_{\omega} \lambda y_1 \dots y_n. P_i\ ,\]
 where possibly $n = 0$;
\end{enumerate}
\item in case of $N \equiv \lambda y_1 \dots  y_n. y_j L_1 \dots
  L_m$, we have
\begin{enumerate}[$-$]
\item $M\equiv \lambda y_1 \dots  y_n. y_j P_1 \dots  P_m$,
\item and for every $i$, with $1 \leq i \leq m$,
\[\lambda y_1 \dots  y_n. L_i =_{\omega}  \lambda y_1 \dots
y_n. P_i\ .
\]
\end{enumerate}
\end{enumerate}
\end{defi}

Recall that a set $\mathcal{X}$ of closed terms, is {\em cofinal}
for $\beta\eta$-reductions, if every closed term $M$ has a
$\beta\eta$-reduct in $\mathcal{X}$.

\begin{defi}
We say that a set $\mathcal{X}$ of closed terms is {\em
supercofinal}  if it is cofinal and contains all the terms that do
not reduce to a zero-term.
\end{defi}

\begin{rem} In the previous Definition, observe that, due to the
cofinality of $\mathcal{X}$, if a term reduces to a zero-term then
it reduces to a zero-term which is in $\mathcal{X}$. \end{rem}

In the following, let $\mathcal{X}$ be a specified
supercofinal set.

\begin{defi}\label{can_end_piece}

An endpiece in standard form
\begin{equation}\label{star7-1}
\eqalign{
F \longrightarrow^{*}_{\beta\eta} H_1 \
_{\beta\eta}^{*}\!&\longleftarrow G_1 M_1 =_{\omega} G_1 N_1
\longrightarrow^{*}_{\beta\eta}
 H_2 \ _{\beta\eta}^{*}\!\longleftarrow\cr
\ _{\beta\eta}^{*}\!&\longleftarrow G_2 M_2 =_{\omega} G_2 N_2 =_{\beta\eta}
\longrightarrow^{*}_{\beta\eta} \dots \qquad\qquad\cr
\dots  \ _{\beta\eta}^{*}\!&\longleftarrow G_t M_t =_{\omega} G_t N_t
\longrightarrow^{*}_{\beta\eta} H_{t+1}
\  _{\beta\eta}^{*}\!\longleftarrow F'
}
\end{equation}
is called {\em an $\mathcal{X}$-canonical endpiece} (or,
when $\mathcal{X}$ is clear from the context, simply {\em a
canonical endpiece}) \iff
\begin{enumerate}[(1)]
\item
for every $i$, $i = 1, \dots , t+1$,
 the confluence terms
$H_i$ belong to $\mathcal{X}$;
\item for every $i$, $i = 1, \dots , t+1$, there exist terms
$Y$, $L_1, \dots , L_m$ $Z_1 \dots 
  Z_n$ (possibly different for different $i$) such that $G_{i}$ has the form
\[G_{i} \equiv \lambda x. \lambda y_1 \dots  y_n. ((\lambda y.Y)L_1 
  \dots  L_m) Z_1 \dots Z_n\ ,
\]
  and such that the following holds:
\begin{enumerate}[(a)]
\item (Conditions on the Left Facing Arrows)\\
for every $i$, $i = 1, \dots , t$, the sequence of left reductions
\[H_i \;_{\beta\eta}^{*}\!\longleftarrow G_{i} M_i\]
has the following structure:
\begin{enumerate}[$-$]
\item a one step $\beta$-reduction of the form
\[[M_{i}/x] (\lambda y_1 \dots  y_n. ((\lambda y.Y)L_1 \dots  L_m) Z_1 \dots 
  Z_n) \:\:\: _{\beta}\longleftarrow G_{i} M_{i}
\]
\item followed by a sequence of non-head
$\beta$-reductions,
\item followed by a sequence of $\eta$-reductions.
\end{enumerate}
\item (Condition on the Right Facing Arrows)\\
for every $i$, $i = 1, \dots , t$, the sequence of right reductions
\[G_{i} N_i\longrightarrow_{\beta\eta}^{*} H_{i+1}\]
has the following structure
\[\eqalign{G_{i} N_i&\longrightarrow_{\beta\eta}^{*}
 [N_{i}/x] (\lambda y_1 \dots  y_n. ((\lambda y.Y)L_1 \dots  L_m) Z_1 \dots 
  Z_n)\cr
&\longrightarrow_{\beta\eta}^{*}  \lambda y_1 \dots  y_n.
[N_{i}/x](\lambda y.Y) [N_{i}/x]L_1 \dots  [N_{i}/x]L_m) y_1 \dots 
  y_n)\cr
&\longrightarrow_{\eta}^{*}  [N_{i}/x](\lambda y.Y)[N_{i}/x]L_1 \dots
               [N_{i}/x]L_m\cr
&\longrightarrow_{\beta\eta}^{*}J_0 J_1 \dots  J_m
\longrightarrow_{\beta\eta}^{*} H_{i+1}\cr}
\]
where
\[ J_0 \equiv  \left \{ \begin{array}{ll}
\mbox{the $\beta\eta$-normal form of $[N_i/x]\lambda y.Y$ } &
\mbox{if exists;}\\
\mbox{$[N_i/x]\lambda y.Y$}  & \mbox{otherwise.}
\end{array} \right. \]
and for $k= 1, \dots , m$
\[ J_k \equiv  \left \{ \begin{array}{ll}
\mbox{the $\beta\eta$-normal form of $[N_i/x]L_k$ } &
\mbox{if exists;}\\
\mbox{$[N_i/x]L_k$}  & \mbox{otherwise.}
\end{array} \right. \]
\end{enumerate}
\end{enumerate}
\end{defi}

In the following definition, recall that an endpiece can be
considered as a finite tree of equality inferences.

\begin{defi} Given the supercofinal set $\mathcal{X}$,
the notion of {\em $\mathcal{X}$-canonical proof}
 is defined inductively as
follows. \begin{enumerate}[$\bullet$]
 \item A $\beta\eta$-conversion is {\em $\mathcal{X}$-canonical} if
the confluence term belongs to $\mathcal{X}$.
\item An instance
of the $\omega$-rule is {\em $\mathcal{X}$-canonical} if the proofs
of the premisses of the instances are $\mathcal{X}$-canonical.
\item Otherwise a proof is {\em canonical}  if its endpiece is an
$\mathcal{X}$-canonical endpiece and all the proofs of the leaves
which are direct conclusions of the $\omega$-rule are
$\mathcal{X}$-canonical.
 \end{enumerate}
\end{defi}
 \begin{prop} For every supercofinal set $\mathcal{X}$,
 every provable equality $M =_{\omega} N$  has an
$\mathcal{X}$-canonical proof.
\end{prop}
\proof Let $\mathcal{X}$ be fixed. We prove this proposition by
induction on the ordinal $\ord(\mathcal{T})$ of a proof $\mathcal{T}$
of $M =_{\omega} N$. For the basis case just suppose that $M
=_{\beta\eta} N$ and use the
Church-Rosser theorem.

      For the induction step we distinguish two cases.

\emph{First Case}. $M =_{\omega} N$ is the direct conclusion of the
$\omega$-rule.
This follows directly from the induction hypothesis.

\emph{Second Case 2}.
 $\mathcal{T}$ has an endpiece of the form
\begin{equation}\label{star7}
\eqalign{
M \longrightarrow^{*}_{\beta\eta} H_1 \
_{\beta\eta}^{*}\!&\longleftarrow G_1 M_1 =_{\omega} G_1 N_1
\longrightarrow^{*}_{\beta\eta}
 H_2 \ _{\beta\eta}^{*}\!\longleftarrow\cr
\ _{\beta\eta}^{*}\!&\longleftarrow G_2 M_2 =_{\omega} G_2 N_2 =_{\beta\eta}
\longrightarrow^{*}_{\beta\eta} \dots \cr
\dots  \ _{\beta\eta}^{*}\!&\longleftarrow G_t M_t =_{\omega} G_t N_t
\longrightarrow^{*}_{\beta\eta} H_{t+1}
\  _{\beta\eta}^{*}\!\longleftarrow N\cr
}
\end{equation}
 where, for each $i = 1 \dots  t$,  $M_i =_{\omega}
N_i$ is the conclusion of an instance of the $\omega$-rule.

Observe that, without changing the ordinal of the proof, we can
assume that every $H_i$, with $1 \leq i \leq t+1$, is in
$\mathcal{X}$.

Consider the first component of the endpiece (\ref{star7})
\[M\longrightarrow_{\beta\eta}^{*} H_1 \
_{\beta\eta}^{*}\!\longleftarrow G_1 M_1 =_{\omega}G_1 N_1
\]
Let $\sigma$ be a standard $\beta\eta$-reduction $G_1 M_1
\longrightarrow_{\beta\eta}^{*} H_1$, with all the $\eta$-reductions
postponed.
We have now different subcases.

{\em First Subcase}. No etrace of $M_1$ appears in functional
position in a head redex neither in the head part of $\sigma$, nor
in $H_1$ itself (that is $H_1$ has not a head redex of the form
$(\lambda x.U)V$, with
$\lambda x. U$ an etrace of $M_1$).

In this case, the same head reductions can be performed (up to a
substitution of $M_1$ by $N_1$) in the $G_1 N_1$ side. Thus simply
replacing $G_1$, we may freely assume that this head part is missing
at all and thus $\sigma$ is composed only of non-head
$\beta$-reductions followed by $\eta$-reductions.
Moreover, by our hypothesis, we can also assume that $G_1$ has not the form:\\
$\lambda x y_1 \ldots y_p.\ x Y_1 \cdots Y_q$.

Moreover we can also assume that $G_1$ begins with a $\lambda$. For
otherwise, assume that in  the head part of $\sigma$, a $\lambda$
never appears at the beginning of the reducts of $G_1$. Therefore
all the reduction $\sigma$ is internal to $G_1$ and $M_1$, and this
implies that $H_1$ has the form $G'_1 M'_1$, where $G_1$
$\beta\eta$-reduces to $G'_1$ and $M_1$ $\beta\eta$-reduces to
$M'_1$, respectively. Thus, replacing $G_1$ with $\lambda x. G'_1
x$, we obtain a term of the required form.

On the $G_1 N_1$ side, the {\em Conditions on the Right Facing
Arrows} may require a reduction of $G_1 N_1$ to a suitable term
$H^{+}$.

By the Church-Rosser Theorem and the cofinality of $\mathcal{X}$,
let $\overline{H}$ be a term in  $\mathcal{X}$, which is a common
reduct of $H^{+}$ and $H_2$. Now, there exists a proof
$\mathcal{T'}$ of $\overline{H} =_{\omega} N$, with
$\ord(\mathcal{T'}) < \ord(\mathcal{T})$ (where $N$ is the final term
of the endpiece (\ref{star7})). Thus by induction hypothesis there
exists a canonical proof $\mathcal{T}_1$ of $\overline{H} =_{\omega}
N$. Now, the required canonical proof is obtained by
concatenating the component
\[M\longrightarrow_{\beta\eta}^{*} H_1 \
_{\beta\eta}^{*}\!\longleftarrow G_1 M_1 =_{\omega}G_1 N_1
\longrightarrow_{\beta\eta}^{*} \overline{H}\ ;\]
with $\mathcal{T}_1$.

That this concatenation results in a canonical proof can be easily
checked in case $\mathcal{T}_1$ ends in an instance of the
$\omega$-rule as well as in case $\mathcal{T}_1$ ends in an
endpiece.\medskip

 {\em Second Subcase}. Assume that:
 \begin{enumerate}[$\bullet$]
 \item an etrace
of $M_1$ appears in functional position in a head redex of the head
part of $\sigma$, or in $H_1$ itself;
\item a $\lambda$
appears at the beginning of some term in  the head part of $\sigma$.
 \end{enumerate}

Thus we have $G_1 M_1 \longrightarrow_{\beta\eta}^{*}
\lambda u.U \longrightarrow_{\beta\eta}^{*} H_1$, for some $U$.
For any closed term $R$, consider the reduction:
              \[G_1 M_1 R \longrightarrow_{\beta\eta}^{*}
              (\lambda u.U)R \longrightarrow_{\beta\eta}
               [R/u]U \longrightarrow_{\beta\eta}^{*} H'\ .\]
Here $H'$ is $[R/u]H_1$. This can be done for every $\lambda$
appearing in the head part of $\sigma$. Thus for each choice of
closed $R_1 \dots  R_n$ we have a standard $\beta\eta$-reduction
$\sigma '$ of $G_1 M_1 R_1 \cdots R_n$ to a term $H''$, which is
$H_1$ with each abstracted variable $u_j$
 substituted by the corresponding closed term $R_j$ (unless this variable has been
 eliminated by $\eta$-reduction: in this case the resulting term is
 applied to $R_j$).

Now, being $\mathcal{X}$ supercofinal, either $H''$ is in
$\mathcal{X}$ or $H''$ $\beta\eta$-reduces to a zero-term $H^0$ in
$\mathcal{X}$, by a reduction $\sigma''$. In this reduction some new
$\lambda$ may appear at the beginning of the term (since we have also
$\eta$-reductions), and we treat this $\lambda$ as before, by applying
all the terms in the reduction some other $R$. Thus we extend the
sequence $R_1 \dots R_n$ to a new sequence $R_1 \dots R_n, R'_1 \dots
R'_m$. Since $H^0$ is a zero-term all the external $\lambda$ appearing
in $\sigma ''$ are eventually eliminated by
$\eta$-reductions. Therefore, starting from $H'' R'_1 \cdots R'_m$ and
applying the reductions in $\sigma ''$, we obtain the term $H^0 R'_1
\cdots R'_m$. Now $H^0 R'_1 \cdots R'_m$ is a zero-term, so that if it
is not in $\mathcal{X}$, the reduction to a suitable term in
$\mathcal{X}$ adds no new $\lambda$s at the beginning of the term. So,
without loss of generality we can assume that $H''$ is in
$\mathcal{X}$, and that $\sigma '$ is a standard $\beta\eta$-reduction
of $G_1 M_1 R_1 \dots R_n$ to $H''$, such that no term in the head
part of $\sigma '$ begins with $\lambda$.

Now in the head reduction part of $\sigma '$, we come to a term $V$
with a head redex of the form: $(\lambda u.W) U$,
where $M_1 \longrightarrow_{\beta\eta}^{*} \lambda u.W$.
Let $V \equiv (\lambda u.W) U U_1 \cdots U_v$, we write $V$ in the
form $(\lambda u.W)[V_1 /x_1 , \dots  ,V_{r_{1}}
/x_{r_{1}}]\overrightarrow{X}_1$, showing all the
etraces $V_1,\ldots, V_{r_{1}}$ of $M_1$ in $V$. Then
$$M_1 [N_1/x_1 , \dots  , N_1/x_{r_{1}}]\overrightarrow{X}_1 =_{\omega}
  N_1 [N_1/x_1 , \dots  , N_1/x_{r_{1}}]\overrightarrow{X}_1\eqno{(*)}$$
has a proof with ordinal (much) less than $\ord(\mathcal{T})$.
Now, consider the component
$$M R_1 \dots  R_n \longrightarrow^{*}_{\beta\eta} H'' \
_{\beta\eta}^{*}\!\longleftarrow M_1 [M_1/x_1 , \dots  ,
M_1/x_{r_{1}}]\overrightarrow{X}_1 =_{\omega}\,  M_1 [N_1/x_1 , \dots  , N_1/x_{r_{1}}]\overrightarrow{X}_1$$
The reduction $M_1 [M_1/x_1 , \dots  ,
M_1/x_{r_{1}}]\overrightarrow{X}_1 \longrightarrow^{*}_{\beta\eta}
H''$ has a head part shorter than $\sigma '$. Thus, iterating the
previous transformation for each occurrence $M_1$ in functional
position in the head reduction part of $\sigma'$, we arrive to a
final sequence of terms $\overrightarrow{X}_s$ such that $M_1
[M_1/x_1 , \dots  , M_1/x_{r_{s}}]\overrightarrow{X}_s$ is the last
such occurrence of $M_1$. Therefore,
for what concerns the component
\[M R_1 \dots  R_n \longrightarrow^{*}_{\beta\eta} H'' \
_{\beta\eta}^{*}\!\longleftarrow M_1 [M_1/x_1 , \dots  ,
M_1/x_{r_{s}}]\overrightarrow{X}_s =_{\omega}
=_{\omega}  M_1 [N_1/x_1 , \dots  , N_1/x_{r_{s}}]\overrightarrow{X}_s\]
we can argue as in the First Subcase above.

On the right hand side, observe that the iteration of the previous
argument gives rise to a chain of equalities (where for simplicity, we
do not consider reduction internal to $M_1$; this does not affect the
argument)
\[\eqalign{
 &N_1 [N_1/x_1 , \dots , N_1/x_{r_{1}}]\overrightarrow{X}_1
  =_{\omega} N R_1 \dots R_n\cr
 &N_1 [N_1/x_1 , \dots ,
  N_1/x_{r_{1}}]\overrightarrow{X}_1 =_{\omega} M_1 [N_1/x_1 , \dots ,
  N_1/x_{r_{1}}]\overrightarrow{X}_1\cr
 &M_1 [N_1/x_1 , \dots ,
  N_1/x_{r_{1}}]\overrightarrow{X}_1 =_{\omega} M_1 [M_1/x_1 , \dots ,
  M_1/x_{r_{1}}]\overrightarrow{X}_1\cr
 &M_1 [M_1/x_1 , \dots ,
  M_1/x_{r_{1}}]\overrightarrow{X}_1 \longrightarrow^{*}_{\beta\eta}
M_1 [M_1/x_1 , \dots , M_1/x_{r_{2}}]\overrightarrow{X}_2\cr
 &M_1
[N_1/x_1 , \dots , N_1/x_{r_{1}}]\overrightarrow{X}_1
\longrightarrow^{*}_{\beta\eta} N_1 [N_1/x_1 , \dots ,
  N_1/x_{r_{2}}]\overrightarrow{X}_2\cr
 &M_1 [M_1/x_1 , \dots ,
  M_1/x_{r_{2}}]\overrightarrow{X}_2 =_{\omega} M_1 [N_1/x_1 , \dots
  ,N_1/x_{r_{2}}]\overrightarrow{X}_2\cr
 &M_1 [N_1/x_1 , \dots ,
  N_1/x_{r_{2}}]\overrightarrow{X}_2 =_{\omega} N_1 [N_1/x_1 , \dots ,
  N_1/x_{r_{2}}]\overrightarrow{X}_2\cr
 & \dots\cr
 &M_1[N_1/x_1 ,
  \dots , N_1/x_{r_{s}}]\overrightarrow{X}_s =_{\omega} N_1 [N_1/x_1 ,
  \dots , N_1/x_{r_{1}}]\overrightarrow{X}_s
  }
\]
  From this chain, by
  Proposition \ref{Fact} of Section \ref{two}, one obtains a proof of
\[M_1 [N_1/x_1 , \dots  , N_1/x_{r_{s}}]\overrightarrow{X}_s
=_{\omega} N R_1 \dots  R_n\ ,
\]
  with an ordinal less than $\ord(\mathcal{T})$. We can also
  substitute $M_1 [N_1/x_1 , \dots ,
    N_1/x_{r_{s}}]\overrightarrow{X}_s$ with a suitable reduct
  $\overline{H}$, meeting both the {\em Conditions on the Right Facing
    Arrows} w.r.t.\ $M_1 [N_1/x_1 , \dots ,
    N_1/x_{r_{s}}]\overrightarrow{X}_s$ and the cofinality condition
  w.r.t.\ $\mathcal{X}$. Still, $\overline{H} =_{\omega} N R_1 \dots
  R_n$ has a proof with ordinal less than $\ord(\mathcal{T})$. Thus by
  induction hypothesis there exists a canonical proof $\mathcal{T}_1$
  of $\overline{H} =_{\omega} N R_1 \dots R_n$.

  Now, we can concatenate the component
\[\eqalign{&M R_1 \dots  R_n
\longrightarrow_{\beta\eta}^{*} H'' \
_{\beta\eta}^{*}\!\longleftarrow (\lambda x. M_1 [x /x_1 , \dots  ,
x/x_{r_{s}}]\overrightarrow{X}_s) M_1 =_{\omega}\cr
&=_{\omega}(\lambda x. M_1 [x/x_1 , \dots  ,
x/x_{r_{s}}]\overrightarrow{X}_s)  N_1
\longrightarrow_{\beta\eta}^{*} \overline{H}\ ;}
\]
with $\mathcal{T}_1$.  That this concatenation results in a canonical
proof can be easily checked in case $\mathcal{T}_1$ ends in an
instance of the $\omega$-rule as well as in case $\mathcal{T}_1$ ends
in an endpiece.

Thus we have proved the following:
\begin{center}
for every $R_1 \dots  R_n$, there exists a canonical proof of
$M R_1 \dots  R_n =_{\omega} N R_1 \dots  R_n$.\end{center}
Now, $n$ applications of the $\omega$-rule give the required
canonical proof of $M =_{\omega} N$.\medskip

 {\em Third Subcase}. \begin{enumerate}[$\bullet$]
 \item an etrace
of $M_1$ appears in functional position in a head redex of the head
part of $\sigma$, or in $H_1$ itself;
\item no $\lambda$
appears at the beginning of some term in  the head part of $\sigma$.
 \end{enumerate}
This case can be treated as the previous one, with the difference
that the resulting canonical proof ends in a canonical endpiece,
rather than in an instance of the $\omega$-rule.\qed

We shall need the following result on $\mathcal{X}$-canonical
proofs.

\begin{prop}\label{application} Let $\mathcal{T}$ be an
$\mathcal{X}$-canonical proof of $M =_{\omega}N$ ending in an
endpiece. Then for every sequence of terms $P_1, \ldots, P_m$, there
exist terms $R_1, \ldots, R_n$ such that the equality $MP_1 \cdots
P_m R_1 \cdots R_n =_{\omega}NP_1 \cdots P_m R_1 \cdots R_n$ has an
$\mathcal{X}$-canonical proof $\mathcal{T}_1$, also ending in an
endpiece, with $rank(\mathcal{T}_1) = rank(\mathcal{T})$.
\end{prop}
\proof Assume that $\mathcal{T}$ has an endpiece of the form:
\begin{equation}\label{star8}\eqalign{
M \longrightarrow^{*}_{\beta\eta} H_1 \
_{\beta\eta}^{*}\!&\longleftarrow G_1 M_1 =_{\omega} G_1 N_1
\longrightarrow^{*}_{\beta\eta}
 H_2 \ _{\beta\eta}^{*}\!\longleftarrow\cr
\ _{\beta\eta}^{*}\!&\longleftarrow G_2 M_2 =_{\omega} G_2 N_2 =_{\beta\eta}
\longrightarrow^{*}_{\beta\eta} \dots \cr
\dots  \ _{\beta\eta}^{*}\!&\longleftarrow G_t M_t =_{\omega} G_t N_t
\longrightarrow^{*}_{\beta\eta} H_{t+1}
\  _{\beta\eta}^{*}\!\longleftarrow N
  }
\end{equation}
 where, for each $i = 1 \dots  t$,  $M_i =_{\omega}
N_i$ is the conclusion of an instance of the $\omega$-rule.

We argue by induction on $t$.
Assume $t = 1$.
Consider the first (and unique) component of the endpiece (\ref{star8})
\[M
\longrightarrow_{\beta\eta}^{*} H_1 \
_{\beta\eta}^{*}\!\longleftarrow G_1 M_1 =_{\omega}G_1 N_1\]
Let $P_1, \ldots, P_m$ be given. We have two cases.

{\em First Case}. $H_1 P_1 \cdots P_m$ is in $\mathcal{X}$. In this
case, the component can directly be transformed into a component of
the right form, using the equality $(\lambda x. (G'_1[ x] P_1 \cdots
P_m))M_1 =_{\omega}((\lambda x. G'_1[ x] P_1 \cdots P_m)) N_1$,
where the applicative context $G'_1[\ ]$ is $G_1 \ [\ ]
$, that is $G_1$ applied to the hole $[\ ]$.

{\em Second Case}. $H_1 P_1 \cdots P_m$ is not in $\mathcal{X}$. In
this case, $H_1 P_1 \cdots P_m$ reduces to a suitable zero-term $H'$
in $\mathcal{X}$. To obtain a component of the right form, we have
to transform $(\lambda x. (G'_1[ x] P_1 \cdots P_m))M_1$ as in the
proof of the previous proposition. This can be done - as shown in
the second subcase of such proof - at the cost (in the worst case)
of applying $(\lambda x. (G'_1[ x] P_1 \cdots P_m)) M_1$ to a
sequence $R_1, \ldots, R_n$ of terms and introducing some additional
leaves each one of
ordinal not greater than the one of $M_1 =_{\omega}N_1$.

Hence the result follows for $t=1$.
Now assume  $t > 1$. Let $P_1, \ldots, P_m$ be given.
By induction hypothesis, for some $R_1, \dots , R_n$ there is a proof
with an endpiece of rank less or equal to $rank(\mathcal{T})$ of
$G_1 N_1 P_1 \cdots P_m R_1 \cdots R_n =_{\omega}NP_1 \cdots P_m R_1
\cdots R_n$.
Now consider the first component of the endpiece (\ref{star8})
\[M
\longrightarrow_{\beta\eta}^{*} H_1 \
_{\beta\eta}^{*}\!\longleftarrow G_1 M_1 =_{\omega}G_1 N_1\]
Again we have two cases.

{\em First Case}. $H_1 P_1 \cdots P_m R_1 \cdots R_n$ is in
$\mathcal{X}$. In this case, the component can directly be
transformed into a component of the right form, using the
equality
\[(\lambda x. (G'_1[ x] P_1 \cdots P_m R_1 \cdots R_n))M_1
=_{\omega}(\lambda x. (G'_1[
x] P_1 \cdots P_m R_1 \cdots R_n)) N_1
\]
 for a suitable applicative context $G'_1[\ ]$.

{\em Second Case}. $H_1 P_1 \cdots P_m R_1 \cdots R_n$ is not in
$\mathcal{X}$. In this case, $H_1 P_1 \cdots P_m R_1 \cdots R_n$
reduces to a suitable zero-term $H'$ in $\mathcal{X}$. To obtain a
component of the right form, we have to transform $(\lambda x. G'_1[
x] P_1 \cdots P_m)M_1$ as in the proof of the previous proposition.
This can be done - as shown in the second subcase of such proof - at
the cost (in the worst case) of applying $(\lambda x. (G'_1[ x] P_1
\cdots P_m R_1 \cdots R_n))M_1$ to a sequence $R'_1, \ldots, R'_k$
of terms and introducing some additional leaves each one of
ordinal not greater than the one of $M_1 =_{\omega}N_1$.

Now again by induction hypothesis there exist $R''_1, \dots , R''_s$
such that there is a proof with an endpiece of rank less or equal to
$rank(\mathcal{T})$ of
\[G_1 N_1 P_1 \cdots P_m R_1 \cdots R_n
R'_1 \cdots R'_k R''_1 \cdots R''_s =_{\omega} NP_1
\cdots P_m R_1
\cdots R_n R'_1 \cdots R'_k R''_1 \cdots R''_s\ .
\]
Observe now that since $H' R'_1 \cdots R'_k$ is a zero-term, we can
obtain a term in $\mathcal{X}$ which is a reduct of $H' R'_1 \cdots
R'_k R''_1 \cdots R''_s$ without introducing new $\lambda$s but
(possibly) only other leaves each one of ordinal not greater than
the one
of $M_1 =_{\omega}N_1$.

So in both cases, the result follows.\qed

 \section{Plotkin Terms}\label{PT}

Recall that $H,M,N,P,Q$ always denote closed terms. Let $\lceil M
\rceil$ denote the Church numeral corresponding to the G\"{o}del
number of the term $M$. We can of course require that any term
occurs infinitely many times (up to $\omega$-equality) in the
enumeration. By Kleene's enumerator construction (\cite{Ba84} 8.1.6)
there exists a combinator $\mathbf{J}$ such
that $\mathbf{J} \lceil M \rceil$ $\beta$-converts to $M$, for every $M$.\\
The combinator $\mathbf{J}$ can be used to enumerate various r.e.
sets of closed terms. In particular, let $\mathcal{X}$ be a r.e. set
of terms, and let $T_\mathcal{X}$ be a term representing the r.e.
function that enumerates $\mathcal{X}$. Set $\mathbf{E} \equiv
\lambda x. \mathbf{J}(T_\mathcal{X} x)$. It is well known that we
can assume that $\mathbf{E}$ is in $\beta\eta$-normal form. We call
$\mathbf{E}$ a {\em generator of} $\mathcal{X}$. As usual we shorten
$\mathbf{E} \underline{n}$ with $\mathbf{E}_{n}$. We also suppress
the dependency
of $\mathbf{E}$ from $\mathbf{J}$ and $\mathcal{X}$, when it is clear from the context.\\
Now, by the methods of proof used in \cite{IS2004}, which make use
of modified forms of the celebrated {\em Plotkin terms} (\cite{Ba84}
17.3.26),
one can prove the following:\\

 \begin{lem}\label{plot1} Given a r.e. set of terms $\mathcal{X}$ and
  a generator $\mathbf{E}$ of $\mathcal{X}$, there exists a term $H$ such that
  for every $M$ the following holds
$$H\mathbf{E}_{0}  =_{\omega} HM\quad\hbox{\iff\ for some $k$,}\quad
 M =_{\omega} \mathbf{E}_{k}\ .\eqno{\qEd}
$$
 \end{lem}

\begin{rem} 
 The Lemma's proof is identical to the proof of Proposition 5 of
 \cite{IS2004}, and consists of two parts:
 \begin{enumerate}[(1)]
 \item to show that if $M =_{\omega} \mathbf{E}_{k}$, for some $k$, then
 $ H\mathbf{E}_{0}  =_{\omega} HM$; this is done by the standard
 argument based on the structure of Plotkin terms;
 \item to show that if $M \neq_{\omega} \mathbf{E}_{k}$, for every $k$, then
 $ H\mathbf{E}_{0}  \neq_{\omega} HM$; this difficult point requires
 a detailed analysis of proofs in $\lambda\omega$, as formulated in
 \cite{IS2004}; this analysis is done in \cite{IS2004}
 and is based on casting such proofs in a
 suitable normal form, called in \cite{IS2004} {\em cascaded
 proofs}.
\end{enumerate}
The proof of the following result has the same structure. We define
suitable Plotkin terms, which makes the "if part" easy to check, and
we rely on the analysis based on {\em cascaded
 proofs} for the "only if part". As the external structure of the
 involved Plotkin terms is the same (zero-terms obtained
 by applying suitable $\beta\eta$-normal forms to
 other $\beta\eta$-normal forms), the proof strictly follows the
 pattern of the proof of Proposition 5 of
 \cite{IS2004} and is omitted.
\end{rem}

\begin{prop}\label{plot4}  There exist two terms $\mathbf{H_1}$ and
$\mathbf{H_2}$ such that
 for every $M$
\[\mathbf{H_1} M  =_{\omega} \mathbf{H_2}\quad\hbox{\iff\ for all $k$,}\quad
 M \neq_{\omega} \underline{k}\ .
\]
 \end{prop}
\proof In \cite{IS2004}, we constructed Plotkin terms $P$ and $Q$
such that
 for every $n$

\[P\underline{n} =_{\omega} Q\underline{n}\quad\hbox{\iff}\quad
 \lower6 pt\hbox to8 cm{%
 \vbox{\noindent $n$ is the G\"{o}del number of a closed term\\
 which does not $\beta\eta$-convert to a Church numeral.}}
\]\smallskip

\noindent Now, let the Plotkin terms $F$ and $G$ be such that
\begin{equation}\label{redF}F_{n} G_{ n} M M_1 M_2   \longrightarrow_{\beta} F_{ n} (F_{
n+1}G_{ n+1}M \<\underline{n}, M, P\underline{n}\> \<\underline{n},
\mathbf{J} \underline{n},
Q\underline{n}\>)\mathbf{\Omega}\mathbf{\Omega}\mathbf{\Omega}
\end{equation}
where $\mathbf{\Omega} \equiv (\lambda x.xx)(\lambda x.xx)$ and, as
usual, the notation $\<X_1, X_2, X_3\>$ stands for Church triple, i.e.
$\<X_1, X_2, X_3\> \equiv \lambda z. zX_1 X_2 X_3$, and
\begin{equation}\label{redG}G_{ n} \longrightarrow_{\beta} F_{ n+1}G_{ n+1} (\mathbf{J}
\underline{n}) \<\underline{n}, \mathbf{J} \underline{n},
Q\underline{n}\>\<\underline{n}, \mathbf{J} \underline{n},
Q\underline{n}\>
\end{equation}
{\bf Claim}. We claim that
for every $M$
\[F_{0} G_{ 0} M M M  =_{\omega}
 F_{0} G_{
   0}\mathbf{\Omega}\mathbf{\Omega}\mathbf{\Omega}\quad\hbox{\iff\ for
   all $k$,}\quad M \neq_{\omega} \underline{k}\ .
\]
To prove the claim, we first show that for every $k$
\begin{equation}\label{plo2000}
\noindent F_{0} G_{
0}\mathbf{\Omega}\mathbf{\Omega}\mathbf{\Omega}=_{\omega}
F_{0}(F_{1} (\dots (F_{k} G_{
k}\mathbf{\Omega}\mathbf{\Omega}\mathbf{\Omega})\dots )\mathbf{\Omega}\mathbf{\Omega}\mathbf{\Omega})
\mathbf{\Omega}\mathbf{\Omega}\mathbf{\Omega}.\end{equation}
Indeed, let $k$ be fixed. Then there exists a $k' \geq 1$, such that
$\mathbf{J} (\underline{k+k'}) =_{\omega} \mathbf{\Omega}$. Then,
since $\mathbf{\Omega} \neq_{\omega} \underline{n}$ for every $n$,
we have that
\[\<\underline{k+k'},\mathbf{\Omega}, P(\underline{k+k'})\>
=_{\omega} \<\underline{k+k'}, \mathbf{J} (\underline{k+k'}),
Q(\underline{k+k'})\>\]
 and (repeatedly applying (\ref{redF})) both terms of
equation (\ref{plo2000}) are $\lambda\omega$-equal to
\[\eqalign{&F_{0}(F_{1} (\dots (F_{k} (F_{k+1} (\dots (F_{k+k'}(
F_{k+k'+1}G_{k+k'+1}\mathbf{\Omega}\<\underline{k+k'},\mathbf{\Omega},
P(\underline{k+k'})\>\cr
&\quad\<\underline{k+k'}, \mathbf{J}
(\underline{k+k'}), Q(\underline{k+k'})\>)
\mathbf{\Omega}\mathbf{\Omega}\mathbf{\Omega})
\dots)\mathbf{\Omega}\mathbf{\Omega}\mathbf{\Omega})
\mathbf{\Omega}\mathbf{\Omega}\mathbf{\Omega})\dots )\mathbf{\Omega}
\mathbf{\Omega}\mathbf{\Omega})
\mathbf{\Omega}\mathbf{\Omega}\mathbf{\Omega}=_{\omega}
\enspace\hbox{(by (\ref{redG}))}\cr
=_{\omega}{}
&F_{0}(F_{1} (\dots (F_{k} (F_{k+1} (\dots (F_{k+k'} G_{
k+k'}\mathbf{\Omega}\mathbf{\Omega}\mathbf{\Omega})\dots )\mathbf{\Omega}\mathbf{\Omega}\mathbf{\Omega})
\mathbf{\Omega}\mathbf{\Omega}\mathbf{\Omega})\dots )\mathbf{\Omega}
\mathbf{\Omega}\mathbf{\Omega})
\mathbf{\Omega}\mathbf{\Omega}\mathbf{\Omega}\ .\cr
}
\]
Assume that $M$ is such that for all $n$,
 $M \neq_{\omega} \underline{n}$.
Let $k$ such that $M =_{\omega} \mathbf{J}\underline{k}$. By
hypothesis, $\<\underline{k},M, P\underline{k}\> =_{\omega}
\<\underline{k}, \mathbf{J} \underline{k}, Q\underline{k}\>$.
It follows that
\[F_{0} G_{ 0} M M M  =_{\omega} F_{0}(F_{1} (\dots (F_{k} G_{
k}\mathbf{\Omega}\mathbf{\Omega}\mathbf{\Omega})\dots )\mathbf{\Omega}\mathbf{\Omega}\mathbf{\Omega})
\mathbf{\Omega}\mathbf{\Omega}\mathbf{\Omega}=_{\omega}  F_{0}
G_{ 0}\mathbf{\Omega}\mathbf{\Omega}\mathbf{\Omega}\ .
\]
Now assume that $M$ is such that for some $n$,
 $M =_{\omega} \underline{n}$. It follows that, for every $k$,
 $\<\underline{k},M, P\underline{k}\> \neq_{\omega} \<\underline{k}, \mathbf{J}
\underline{k}, Q\underline{k}\>$ and, roughly speaking, in the term $
F_{0} G_{ 0} M M M$, $M$ can never be eliminated by $G$. A formal
proof argues by contradiction on a {\em cascaded proof} of
\[F_{0}
G_{ 0} M M M  =_{\omega}
 F_{0} G_{ 0}\mathbf{\Omega}\mathbf{\Omega}\mathbf{\Omega}\ .\]
This ends the proof of the claim. 

Now define
$$\mathbf{H_1} \equiv \lambda x. F_{0} G_{ 0} xxx\quad\hbox{and}\quad
\mathbf{H_2}\equiv F_{0} G_{
0}\mathbf{\Omega}\mathbf{\Omega}\mathbf{\Omega}\ .\eqno{\qEd}$$\medskip

 We shall make extensive use of terms $\mathbf{H_1}$ and
$\mathbf{H_2}$ in the following Section.

\section{Barendregt Construction}\label{BarConstr}

In the present Section, we shall make use of the Proposition 4 of
\cite{IS2004}, that we restate here for the sake of the reader.\medskip

{\em If $M =_{\omega} N$ and $M$ has a $\beta\eta$-normal form then
$N$ has the same normal form. Therefore two $\beta\eta$-normal forms
equalized in
$\lambda\beta\omega$ are identical.}\medskip

We make the following definitions, which will hold in all the present
and the next Section:

\begin{defi}\hfill
\begin{enumerate}[(1)]
\item $\mathbf{\Theta} \equiv (\lambda ab. b(aab)) (\lambda ab.
b(aab))$ (Turing's fixed point).

\item $\mathbf{W} \equiv \lambda xy. xyy$

\item $\mathbf{L} \equiv (\lambda xyz. \lambda abc. xy(z(yc))bac)$

\item $F \equiv \mathbf{\Theta}\mathbf{L}\mathbf{H_1} \equiv  (\lambda ab.b(aab))(\lambda
ab.b(aab))(\lambda xyz. \lambda abc. xy(z(yc))bac) \mathbf{H_1}$

\item $G \equiv \mathbf{\Theta}\mathbf{W}\mathbf{H_2} \equiv  (\lambda ab.b(aab))(\lambda ab.b(aab))
(\lambda xy. xyy) \mathbf{H_2}$

\end{enumerate}
\end{defi}\medskip

Observe:
\[\eqalign{\hbox{(i)}\enspace G 
&\longrightarrow_{\beta\eta} (\lambda b. b(\mathbf{\Theta}b))
  \mathbf{W} \mathbf{H_2}\cr
&\longrightarrow_{\beta\eta}
  \mathbf{W}(\mathbf{\Theta}\mathbf{W})\mathbf{H_2}\cr
&\longrightarrow_{\beta\eta}\mathbf{\Theta}\mathbf{W}
  \mathbf{H_2} \mathbf{H_2} \equiv G \mathbf{H_2}\cr\cr\cr\cr\cr\cr\cr
  }
  \eqalign{\hbox{(ii)}\enspace F Z A B C 
&\longrightarrow_{\beta\eta}  (\lambda b.b(\mathbf{\Theta}b))
  \mathbf{L} \mathbf{H_1} Z A B C \cr
&\longrightarrow_{\beta\eta}\mathbf{L}(\mathbf{\Theta}\mathbf{L}) 
  \mathbf{H_1} Z A B C\cr
&\longrightarrow_{\beta\eta} (\lambda yz. \lambda abc.
  \mathbf{\Theta}\mathbf{L}y(z(yc))bac) \mathbf{H_1} Z A B C\cr
&\longrightarrow_{\beta\eta}(\lambda z. \lambda abc. 
  \mathbf{\Theta}\mathbf{L}\mathbf{H_1}(z(\mathbf{H_1}c))bac) Z A B C\cr
&\longrightarrow_{\beta\eta}(\lambda abc. \mathbf{\Theta}\mathbf{L}
  \mathbf{H_1}(Z(\mathbf{H_1}c))bac) A B C\cr
&\longrightarrow_{\beta\eta}(\lambda bc. \mathbf{\Theta}\mathbf{L}
  \mathbf{H_1}(Z(\mathbf{H_1}c))bAc) B C\cr
&\longrightarrow_{\beta\eta}(\lambda c. \mathbf{\Theta}\mathbf{L}
  \mathbf{H_1}(Z(\mathbf{H_1}c))BAc) C\cr
&\longrightarrow_{\beta\eta}\mathbf{\Theta}\mathbf{L}
  \mathbf{H_1}(Z(\mathbf{H_1}C))BAC \equiv F Z^{*} B A C\cr
 \rlap{where we have shortened $Z(\mathbf{H_1}C)$ to $Z^{*}$.}
  \phantom{F Z A B C}
  }
\]\medskip

\noindent Let $gk$ be the cofinal Gross-Knuth strategy defined in
\cite{Ba84} 13.2.7. By writing $gk(M)$, we mean the term obtained by
starting with the term $M$ and applying (once) the $gk$ strategy.

Then the reduction sequences
\begin{equation}\label{coG}
\eqalign{G
&\longrightarrow_{\beta\eta}^{*} G (gk(\mathbf{H_2}))
 \longrightarrow_{\beta\eta}^{*} G \mathbf{H_2} (gk(\mathbf{H_2}))
 \longrightarrow_{\beta\eta}^{*}\cr
&\longrightarrow_{\beta\eta}^{*}  G (gk(\mathbf{H_2}))
 (gk(gk(\mathbf{H_2}))))
 \longrightarrow_{\beta\eta}^{*} \cdots\cr
 }
\end{equation}

\begin{equation}\label{coF}
\eqalign{F Z A B C
&\longrightarrow_{\beta\eta}^{*} F (gk(Z^*)) (gk(B)) (gk(A)) (gk(C))
 \longrightarrow_{\beta\eta}^{*}\cr
&\longrightarrow_{\beta\eta}^{*}  F (gk((gk(Z^*))^{*}))
 (gk(gk(A))) (gk(gk(B))) (gk(gk(C))) \longrightarrow_{\beta\eta}^{*}
\cdots\cr
 }
\end{equation}
(where again the notation $X^{*}$ is a shortening of $X(\mathbf{H_1}C)$)
are cofinal for $\beta\eta$-reductions starting with $G$
and, respectively, with $F Z A B C$.

Let $P$ be the initial term or an intermediate term of the reduction
sequence of the form (\ref{coG}), we indicate by $\overline{gk}(P)$
the first term in the sequence, which has the form displayed in
(\ref{coG}), obtained from $P$ by the reductions
in (\ref{coG}).

Similarly, if $P$ is the initial term or an intermediate term of a
reduction sequence of the form (\ref{coF}), starting from $F M_1 M_2
M_3 M_4$, for some $M_1, M_2, M_3, M_4$, we indicate by
$\overline{gk}(P)$ the first term in the sequence, which has the
form displayed in (\ref{coF}), obtained from $P$ by the reductions
in (\ref{coF}).

Now, we choose the cofinal set $\mathcal{X}$ as follows:
for every closed term $M$,
\begin{enumerate}[$\bullet$]
\item if $M$ $\beta\eta$-reduces, by the leftmost outermost reduction
strategy,
 to a term of the form $P N_1 \cdots N_k$, where $P$ is a term of
the sequence (\ref{coG}) then $\overline{gk}(P)gk(N_1) \cdots
gk(N_k)$ is the reduct of $M$ in $\mathcal{X}$;
\item if $M$ $\beta\eta$-reduces, by the leftmost outermost reduction
strategy,
 to a term the form $P N_1 \cdots N_k$,
where $P$ is a term of a sequence (\ref{coF}), starting from $F M_1
M_2 M_3 M_4$, for some $M_1, M_2, M_3, M_4$,
 then
\[\overline{gk}(F M_1 M_2 M_3 M_4)gk(N_1) \cdots gk(N_k)\] is the
reduct of $M$ in $\mathcal{X}$;
\item $M$ is in
$\mathcal{X}$, otherwise.
\end{enumerate}

Observe that we use the leftmost outermost reduction strategy, since
it is cofinal (see \cite{Ba84}, 13.1.3). The following Lemma is
immediate.
\begin{lem}
$\mathcal{X}$ is supercofinal.\qed
\end{lem}

\begin{lem}\label{cof1} If $GM_1 \dots  M_m =_{\omega} GN_1 \dots
  N_m$ then, for each $k$, $1 \leq k \leq m$, $M_k =_{\omega} N_k$.
\end{lem}
\proof By induction on the ordinal of a canonical proof
$\mathcal{T}$ of $GM_1 \dots  M_m =_{\omega} GN_1 \dots  N_m$.

{\em Basis}: $\ord(\mathcal{T})$ is $1$. This case is clear since $G$
is of order $0$.

{\em Induction step}:

\noindent{\em Case 1}. $\mathcal{T}$ ends in an application of the
$\omega$-rule.
 Apply the induction
hypothesis to the subproof of
$G M_1 \dots  M_m \mathbf{I} =_{\omega}G N_1 \dots  N_m \mathbf{I}$.

\noindent{\em Case 2}. $\mathcal{T}$ has the endpiece 
\[\eqalign{G M_1 \dots  M_m
 &\longrightarrow_{\beta\eta}^{*} R_1 \; _{\beta\eta}^{*}\!\longleftarrow
 L_1P_1 =_{\omega}L_1Q_1
  \longrightarrow_{\beta\eta}^{*} R_2 \; _{\beta\eta}^{*}\!\longleftarrow
    L_2P_2 =_{\omega}L_2 Q_2 \cr
 &\longrightarrow_{\beta\eta}^{*}\dots
  \longrightarrow_{\beta\eta}^{*}  R_{t+1} \; _{\beta\eta}^{*}\!\longleftarrow
 G N_1 \dots  N_m\ .
  }
\]
Since $\mathcal{T}$ is canonical, every term $R_i$, with $1 \leq i
\leq t+1$, has the form
\[\mathbf{\Theta} \mathbf{W} H_{i,1}^{*} \dots  H_{i, n_{i}}^{*}
M_{i,1}^{*} \dots  M_{i,m}^{*}
\]
where $H_{i,j}^{*} =_{\omega} \mathbf{H_2}$, for $j=1, \dots  ,n_i$,
and $M_{i,k}^{*} =_{\omega} M_k$ for $ k = 1, \dots  ,m$. Since $G$ is
of order $0$, we must also have $M_{t+1,k}^{*} =_{\omega}
N_k$ for $k=1, \dots  ,m$. This completes the proof.\qed

By an inspection of the proof of the previous Lemma, the following
stronger result can be obtained.

\begin{lem}\label{cof11} Assume that $GM_1 \dots  M_m =_{\omega} GN_1 \dots  N_m$
has a canonical proof $\mathcal{T}$. Then for each $k$, $1 \leq k
\leq m$, there is a canonical proof $\mathcal{T}_k$ of $M_k
=_{\omega} N_k$, with the ordinal of $\mathcal{T}_k$ not greater
than the ordinal of $\mathcal{T}$.\qed
\end{lem}

For the proof of the following Lemma, we need Proposition 4 of
\cite{IS2004}, stated above.

\begin{lem}\label{cof2} Suppose that:
\begin{enumerate}[$\bullet$]
\item $F L_1 P_1 Q_1 \underline{n} M_1 \dots  M_m =_{\omega} F L_2 P_2 Q_2
\underline{n} N_1 \dots  N_m$;

\item $L_1 =_{\omega} G(\mathbf{H_1}\underline{n}) \dots 
 (\mathbf{H_1}\underline{n})$, $k$ times;

\item $L_2 =_{\omega} G(\mathbf{H_1}\underline{n}) \dots 
 (\mathbf{H_1}\underline{n})$, $l$ times.
\end{enumerate}
      Then either $P_1 =_{\omega} P_2, Q_1 =_{\omega} Q_2$ and
     $k = l \  mod \ 2$ or $P_1 =_{\omega} Q_2$, $Q_1 =_{\omega} P_2$ and
     $k = l+1 \ mod 2$,
where, possibly, $k = 0$ or $l = 0$.
\end{lem}
\proof

By induction on the ordinal of a canonical
proof of
\[F L_1 P_1 Q_1 \underline{n} M_1 \dots  M_m =_{\omega}
    F L_2 P_2 Q_2 \underline{n} N_1 \dots  N_m
\]

{\em Basis}: The ordinal is $1$ and we have a
$\beta\eta$-conversion. Use a standard argument, taking into account
that by Proposition \ref{plot4} the copies of $\mathbf{H_2}$ are
distinct, w.r.t. $\omega$-equality, from the copies of
$\mathbf{H_1}\underline{n}$. Therefore the $\beta$-reduction of $G$
cannot affect the count of the copies of $\mathbf{H_1}\underline{n}$.\\

{\em Induction step}:

\noindent{\em Case 1}. The proof ends in an application of the $\omega$-rule.
Just apply the induction
hypothesis to any of the premises.

\noindent{\em Case 2}. The proof has a canonical endpiece beginning with a
component
\[F L_1 P_1 Q_1 \underline{n} M_1 \dots  M_m
\longrightarrow_{\beta\eta}^{*}
 H \; _{\beta\eta}^{*}\!\longleftarrow LQ =_{\omega}LR \longrightarrow_{\beta\eta}^{*}
  H^{+}\ .
\]
Now $H$ has the same form as $F L_1 P_1 Q_1 \underline{n}
M_1 \dots  M_m$ by the choice of the cofinal set. W.l.o.g. we can
assume that the reduction from $F L_1 P_1 Q_1 \underline{n} M_1 \dots 
M_m$ to $H$ is a standard $\beta$-reduction followed by a sequence
of $\eta$-reductions. The 8 term head reduction cycle of $F$ with 4
arguments must be completed an integral number of times to result in
a term which $\eta$-reduces to one with $F$
at the head. Suppose that this cycle is completed $s$ times. Let $r =
k+s$.

On the other hand, since the endpiece is canonical $L$, after a
sequence
(possibly empty) of $\eta$-reductions, reduces to a term of the form
\[\lambda z.  X_0  X_2  X_3  X_4  X_5  X_6  X_7  X_8  Y_1 \dots   Y_m\]
where  $X_0 \equiv \lambda x.  X_1$.
Indeed, the form of the external structure of $H$ must be
\[X'_0 X'_2
X'_3 X'_4  X'_5  X'_6  X'_7  X'_8  Y'_1 \dots  Y'_m\ ,
\]
 since this is
the form of any term, in the cofinal sequence, starting from $F L_1
P_1 Q_1 \underline{n} M_1 \dots  M_m$. Therefore $L$ must have the
form
\[\lambda z. X_0 X_2 X_3 X_4 X_5 X_6 X_7 X_8 Y_1 \dots  Y_m\ ,\]
since we have to obtain $H$ by internal reductions and $Q$ is not
substituted
for a variable in functional position in a head redex.
It follows, using for some items Proposition 4 of \cite{IS2004},
that:
\begin{enumerate}[$\bullet$]
\item $[Q/z]X_0
\longrightarrow_{\beta\eta}^{*} \lambda ab. b(aab)$ (since $\lambda
ab. b(aab)$ is in $\beta\eta$-normal form);
\item $ [Q/z]  X_2
\longrightarrow_{\beta\eta}^{*} \lambda ab. b(aab);$
\item $ [Q/z]  X_3
\longrightarrow_{\beta\eta}^{*} \mathbf{L}$ (since $\mathbf{L}$ is
in $\beta\eta$-normal form);
\item $[Q/z]X_4 =_{\omega} \mathbf{H_1}$
\item $[Q/z]X_5
=_{\omega} G\mathbf{H_2} \dots  \mathbf{H_2}(\mathbf{H_1}\underline{n})
\dots (\mathbf{H_1}\underline{n})$,\\ with $t$ occurrences of
$\mathbf{H_2}$, due to the possible $\beta$-reduction of $G$, and
$r$ occurrences of $\mathbf{H_1}\underline{n}$, since we have
started with $k$ copies of $\mathbf{H_1}\underline{n}$, and each
reduction cycle of $F$ adds a copy.
\item $[Q/z]X_6 =_{\omega} P_1$ if $s \equiv 0 \  mod \ 2$ or \\
$[Q/z]X_6 =_{\omega} Q_1$ if $s \equiv 1 \ mod \ 2$,\\
this item, and the following one, results from the fact the each
reduction cycle of $F$ interchanges $P_1$ and $Q_1$;
\item $[Q/z] X_7 =_{\omega} Q_1$ if $s \equiv 0 \ mod \ 2$ or \\
$[Q/z] X_7 =_{\omega} P_1$ if $s \equiv 1 \ mod \ 2$;
\item $[Q/z] X_8 \longrightarrow_{\beta\eta}^{*} \underline{n}$;
\item $ [Q/z] Y_i =_{\omega} M_i$, for every $1 \leq i \leq m$.
\end{enumerate}

From the fact that $P =_{\omega} R$, and using again Proposition 4
of \cite{IS2004}, we have:
\begin{enumerate}[$\bullet$]
\item
  $[R/z] X_0 \longrightarrow_{\beta\eta}^{*}
\lambda ab. b(aab)$;
\item $ [R/z]  X_2 \longrightarrow_{\beta\eta}^{*} \lambda
ab. b(aab)$;
\item $ [R/z]  X_3 \longrightarrow_{\beta\eta}^{*} \mathbf{L}$;
\item $ [R/z]  X_4
=_{\omega} \mathbf{H_1}$;
\item
$ [R/z]  X_5 =_{\omega} G\mathbf{H_2} \dots 
\mathbf{H_2}(\mathbf{H_1}\underline{n})
\dots (\mathbf{H_1}\underline{n})$ with $t$ occurrences of
$\mathbf{H_2}$ and $r$ occurrences of $\mathbf{H_1}\underline{n}$;
\item
$[R/z]  X_6 =_{\omega} P_1$ if $s \equiv 0 \ mod \ 2$;
\item $[R/z]  X_6 =_{\omega} Q_1$ if $s \equiv 1 \ mod \ 2$;
\item $[R/z]  X_7 =_{\omega} Q_1$ if $s \equiv 0 \ mod \ 2$;
\item $[R/z]  X_7 =_{\omega} P_1$ if $s \equiv 1 \ mod \ 2$;
\item $[R/z]  X_8 \longrightarrow_{\beta\eta}^{*} \underline{n}$;
\item $[R/z] Y_i
=_{\omega} M_i$ , for every $1 \leq i \leq m$.
\end{enumerate}

Observe moreover that $H^{+}$, because of its construction, has the
same form as $H$ (up to some $\eta$-reductions). Say $H^{+} \equiv F
L_{1}^{+} P_{1}^{+} Q_{1}^{+} \underline{n} M_{1}^{+} \dots 
M_{m}^{+}$. Moreover, we can freely assume that $L_{1}^{+}$ is
$G(\mathbf{H_1}\underline{n}) \dots (\mathbf{H_1}\underline{n})$ with
no occurrence of $\mathbf{H_2}$ and $r$ occurrences of
$\mathbf{H_1}\underline{n}$. This amounts to start with a different
term and then perform $t$ $\beta$-reductions of $G$. By Proposition
\ref{plot4} the copies of $\mathbf{H_2}$ are distinct, w.r.t.
$\omega$-equality, from the copies of $\mathbf{H_1}\underline{n}$.
Therefore the $\beta$-reduction of $G$
cannot affect the count of the copies of $\mathbf{H_1}\underline{n}$.

The part of the proof beginning with $H^{+}$ is a canonical proof of
the fact that $H^{+} =_{\omega} F L_2 P_2 Q_2 \underline{n} N_1 \dots
N_m$, because the cofinality restriction met for $LR$ also works for
$H^{+}$ . Thus the induction hypothesis applies to this proof.

Now the idea is that $r$ and $l$ have "to be in accordance" by
induction hypothesis. On the other hand $k$ differs from $r$ only for
$s$ cycles of $F$, and therefore they behave in the right way.  So the
required property is obtained by transitivity. Formally:\medskip

{\em Subcase 2.1}.  $P_{1}^{+} =_{\omega} P_2$, $Q_{1}^{+} =_{\omega}
Q_2$ and $r \equiv l \ mod \ 2$.\\ In case $s$ is even we have $P_1
=_{\omega} P_2$ and $Q_1 =_{\omega} Q_2$ and $k \equiv l \ mod
\ 2$. In case $s$ is odd we have $k$ and $l$ with opposite parity and
$Q_1 =_{\omega} P_{1}^{+} =_{\omega} P_2$, $P_1 =_{\omega} Q_{2}^{+}
=_{\omega} Q_2$.\medskip

{\em Subcase 2.2}. $P_{1}^{+} =_{\omega} Q_2$, $Q_{1}^{+} =_{\omega}
P_2$ and
$r \equiv l+1 \ mod \ 2$.\\
In case $s$ is even we have $P_1 =_{\omega} Q_2$ and $Q_1 =_{\omega}
P_2$ and $k \equiv l+1 \ mod \ 2$. In case $s$ is odd we have $k$
and $l$ with the same parity and $P_1 =_{\omega} Q_{1}^{+}
=_{\omega} P_2$, $Q_1 =_{\omega} P_{2}^{+} =_{\omega} Q_2$. This
completes the proof.   \qed

Also in this case, by an inspection of the proof, the following
stronger result can be obtained.

\begin{lem}\label{cof22}
Suppose that
\begin{enumerate}[$\bullet$]
\item $F L_1 P_1 Q_1 \underline{n} M_1 \dots  M_m =_{\omega} F L_2 P_2 Q_2
\underline{n} N_1 \dots  N_m$ has a canonical proof $\mathcal{T}$;

\item $L_1 =_{\omega} G(\mathbf{H_1}\underline{n}) \dots 
 (\mathbf{H_1}\underline{n})$, $k$ times, has a canonical proof $\mathcal{T}_1$;

\item $L_2 =_{\omega} G(\mathbf{H_1}\underline{n}) \dots 
 (\mathbf{H_1}\underline{n})$, $l$ times, has a canonical proof $\mathcal{T}_2$.
\end{enumerate}
      Then:
      \begin{enumerate}[$\bullet$]
      \item either  $k = l \  mod \ 2$ and $P_1 =_{\omega} P_2, Q_1 =_{\omega} Q_2$
       have canonical proofs $\mathcal{T}_3$ and, respectively,
       $\mathcal{T}_4$,
 \item or $k = l+1 \ mod \ 2$ and $P_1 =_{\omega} Q_2$, $Q_1 =_{\omega} P_2$ have canonical proofs $\mathcal{T}_3$ and, respectively,
       $\mathcal{T}_4$.
       \end{enumerate}
\noindent Here, possibly, $k = 0$ or $l = 0$, and
$\max{\{\ord(\mathcal{T}_3),\ord(\mathcal{T}_4)\}} \leq
\max{\{\ord(\mathcal{T}),\ord(\mathcal{T}_1),\ord(\mathcal{T}_2)\}}$.\qed
\end{lem}

\subsection{Well Founded Trees}
We assume that we have encoded sequences of numbers as numbers, with
$0$ encoding the empty sequence. $\<n\>$ is the sequence consisting of
$n$ alone (singleton) and $*$ is the concatenation function. For
simplicity, we shall use these notations ambiguously for the
corresponding $\lambda$-terms. We require only that the term $y\ * \
\<z\>$ is in $\beta\eta$-normal form with $z\mathbf{I}\mathbf{I}$ at
its head (this construction can be obtained
"making normal" a term representing $*$, see \cite{St2}).

Our proof of  $\mathbf{\Pi_{1}^{1}}$-completeness of $\lambda\omega$
is inspired by the argument in Section 17.4 of \cite{Ba84} (however,
we will substantially modify Barendregt's construction). The
starting point is the following well known theorem (see
\cite{Rogers} Ch.16 Th.20):

\begin{thm}
The set of (indices of) well-founded recursive trees is
$\mathbf{\Pi_{1}^{1}}$-complete.\qed
\end{thm}

The idea is now to reduce the well-foundedness of a recursive tree to
the equality of two suitable terms in $\lambda\omega$.

Suppose that we have a primitive recursive tree $\mathbf{t}$ with a
representing term $\mathbf{T}$ such that
\[ \mathbf{T}\underline{n} \longrightarrow_{\beta\eta}^{*} \left \{ \begin{array}{ll}
\mathbf{I} & \mbox{if $n$ is the number of a sequence in $\mathbf{t}$;}\\
\mathbf{K}^{*} &  \mbox{otherwise.}
\end{array} \right. \]
Define
\[\eqalign{A \equiv_{def} 
 &\mathbf{\Theta}(\lambda x.\lambda a. a
          (\lambda y.\mathbf{T}y(\lambda z. FG(x\mathbf{K}(y\,*\,\<z\>))\cr
 &(x\mathbf{K}^{*}(y\,*\,\<z\>))z))
          (\lambda y.\mathbf{T}y(\lambda z. FG(x\mathbf{K}^{*}(y\,*\,\<z\>))\cr
 &(x\mathbf{K}(y\,*\,\<z\>))z)))\mathbf{K}\cr
  B \equiv_{def}
&\mathbf{\Theta}(\lambda x.\lambda a. a
          (\lambda y.\mathbf{T}y(\lambda z. FG(x\mathbf{K}(y\,*\,\<z\>))\cr
&(x\mathbf{K}^{*}(y\,*\,\<z\>))z))
          (\lambda y.\mathbf{T}y(\lambda z. FG(x\mathbf{K}^{*}(y\,*\,\<z\>))\cr
&(x\mathbf{K}(y\,*\,\<z\>))z)))\mathbf{K}^{*}\cr
  }
\]
Clearly:
\[\eqalign{A
&\longrightarrow_{\beta\eta}^{*}  \lambda y.\mathbf{T}y(\lambda z.
  FG(A(y\,*\,\<z\>))(B(y\,*\,\<z\>))z)\cr
B
&\longrightarrow_{\beta\eta}^{*}  \lambda y.\mathbf{T}y(\lambda z.
  FG(B(y\,*\,\<z\>))(A(y\,*\,\<z\>))z)\cr
  }
\]

Now we state a corollary to Lemma \ref{cof22}.
\begin{cor}\label{41}
If $FG(A\underline{n})(B\underline{n})\underline{n}M_1 \dots  M_m
=_{\omega} FG(B\underline{n})(A\underline{n})\underline{n}N_1 \dots 
N_m$ has a canonical proof $\mathcal{T}$ then $A\underline{n}
=_{\omega} B\underline{n}$ has a canonical proof $\mathcal{T}_1$,
with $\ord(\mathcal{T}_1)\leq \ord(\mathcal{T})$.\qed
\end{cor}

\begin{lem}\label{4} If the subtree $\mathbf{t}(n)$
of the tree $\mathbf{t}$ rooted at $n$ is well-founded
          then $A\underline{n} =_{\omega} B\underline{n}$.
\end{lem}
\proof By induction on the ordinal of the subtree $\mathbf{t}(n)$,
which is defined in the natural way. Note that if $n$ is not the
number of a sequence in the tree then
 $ \mathbf{T}\underline{n} \longrightarrow_{\beta\eta}^{*} \mathbf{K}^{*}$ so
 $A\underline{n} \longrightarrow_{\beta\eta}^{*} \mathbf{I}
  \; _{\beta\eta}^{*}\!\longleftarrow B\underline{n}$.

{\em Basis}. The ordinal is $0$ so the tree $\mathbf{t}(n)$ contains
only the empty sequence. Suppose that $0$ is the number of the empty
sequence. Then
\[A\underline{0} \longrightarrow_{\beta\eta}^{*}
 \lambda z. FG(A(\underline{0}\,*\,\<z\>))(B(\underline{0}\ * \
 \<z\>))z
\]
 and
\[B\underline{0} \longrightarrow_{\beta\eta}^{*} \lambda z.
FG(B(\underline{0}\,*\,\<z\>))(A(\underline{0}\,*\,\<z\>))z
\]
and if $N$ $\beta\eta$-converts to a Church numeral then
\[A\underline{0}N \longrightarrow_{\beta\eta}^{*}
FG\mathbf{I}\mathbf{I}N
\; _{\beta\eta}^{*}\!\longleftarrow B\underline{0}N
\]
and if $N$ does not
$\beta\eta$-convert to a  Church numeral then
\[\eqalign{A\underline{0}N 
  \longrightarrow_{\beta\eta}^{*}{}& FG(A(\underline{0}\,*\,\<N\>))
  (B(\underline{0}\,*\,\<N\>))N\longrightarrow_{\beta\eta}^{*}\cr
  \longrightarrow_{\beta\eta}^{*}{}& F(G(\mathbf{H_1} N))
  (B(\underline{0}\,*\,\<N\>))(A(\underline{0}\,*\,\<N\>))N
  =_{\omega}\cr
  =_{\omega}{}& F(G\mathbf{H_2})(B(\underline{0}\ * \
  \<N\>))(A(\underline{0}\,*\,\<N\>))N
  \; _{\beta\eta}^{*}\!\longleftarrow B\underline{0}N\ .\cr
  }
\]
So, by the $\omega$-rule, $A\underline{0} =_{\omega}B\underline{0}$.\smallskip

{\em   Induction Step}. The
ordinal of the subtree rooted at $n$ is larger than $0$. We have
\[\eqalign{A\underline{n} 
 &\longrightarrow_{\beta\eta}^{*} \lambda z.
  FG(A(\underline{n}\,*\,\<z\>))(B(\underline{n}\,*\,\<z\>))z\cr
  B\underline{n} 
 &\longrightarrow_{\beta\eta}^{*} \lambda z.
  FG(B(\underline{n}\,*\,\<z\>))(A(\underline{n}\,*\,\<z\>))z\cr
  }
\]
Now, if $N$ $\beta\eta$-converts
to a Church numeral, then
\[\eqalign{A\underline{n}N 
  \longrightarrow_{\beta\eta}^{*}{}&
  FG(A(\underline{n}\,*\,\<N\>))(B(\underline{n}\,*\,\<N\>))N =_{\omega}
  \quad\hbox{(by induction hypothesis)}\cr
  =_{\omega}{}& FG(B(\underline{n}\,*\,\<N\>))(A(\underline{n}\ * \
  \<N\>))N \; _{\beta\eta}^{*}\!\longleftarrow B\underline{n}N\cr
  }
\]
and if $N$ does not $\beta\eta$-convert to a  Church
numeral, then
\[\eqalign{A\underline{n}N 
 \longrightarrow_{\beta\eta}^{*}{}&
  FG(A(\underline{n}\,*\,\<N\>))(B(\underline{n}\,*\,\<N\>))N
  \longrightarrow_{\beta\eta}^{*}\cr
 \longrightarrow_{\beta\eta}^{*}{}& F(G(\mathbf{H_1} N))
  (B(\underline{n}\,*\,\<N\>))(A(\underline{n}\,*\,\<N\>))N
 =_{\omega}\cr
 =_{\omega}{}& F(G\mathbf{H_2})(B(\underline{n}\
  * \ \<N\>))(A(\underline{n}\,*\,\<N\>))N\; _{\beta\eta}^{*}\!\longleftarrow
  B\underline{n}N\ .\cr
  }
\]
So by the $\omega$-rule $A\underline{n} =_{\omega} B\underline{n}$.
This completes the proof.\qed

\begin{lem}\label{semifinal} If $A\underline{n} =_{\omega} B\underline{n}$
then the subtree $\mathbf{t}(n)$ rooted at $n$ is well-founded or
$n$ is not in the tree $\mathbf{t}$.
\end{lem}
\proof
 Consider all canonical proofs of smallest ordinal of
$A\underline{n} =_{\omega} B\underline{n}$ for $n$ in the tree
$\mathbf{t}$, and assume that the subtree $\mathbf{t}(n)$
rooted at $n$ is not well-founded. Let $\mathcal{T}$ be such a proof.

{\em Case 1}. $\mathcal{T}$ is a $\beta\eta$-conversion. It is
easily seen that this is impossible. Indeed, assume that
$A\underline{n} =_{\beta\eta} B\underline{n}$; by the Church-Rosser
Theorem
a common $\beta\eta$-reduct must exist.

On the other hand, since $n$ is in $\mathbf{t}$, we have
\[A\underline{n} \longrightarrow_{\beta\eta}^{*}
 \lambda z. FG(A(\underline{n}\,*\,\<z\>))(B(\underline{n}\ * \
 \<z\>))z
\]
 and
\[B\underline{n} \longrightarrow_{\beta\eta}^{*} \lambda z.
FG(B(\underline{n}\,*\,\<z\>))(A(\underline{n}\,*\,\<z\>))z\ .
\]
However that $\lambda
z. FG(A(\underline{n}\,*\,\<z\>))(B(\underline{n}\,*\,\<z\>))z$ and
$\lambda z. FG(A(\underline{n}\,*\,\<z\>))(B(\underline{n}\,*\,
\<z\>))z$ have a common reduct is impossible, considering that
$A(\underline{n}\,*\,\<z\>)$ and $B(\underline{n}\,*\, \<z\>)$ are not
$\beta\eta$-convertible and at each reduction step of $F$ they are
interchanged and a new term $\mathbf{H_1}$ is generated. So the
reducts never have the right "parity" to be identical (see also Lemma
\ref{cof2}).

{\em Case 2}. $\mathcal{T}$ ends in the $\omega$-rule. Then for each
$m$, $A \underline{n} \underline{m} =_{\omega}
 B \underline{n} \underline{m}$ has a canonical proof of smaller ordinal. Now
\[\eqalign{A \underline{n} \underline{m} 
 &\longrightarrow_{\beta\eta}^{*}
  \lambda y.\mathbf{T}y(\lambda z. FG(A(y\,*\,\<z\>))(B(y\,*\,\<z\>))z)
  \underline{n} \underline{m}\cr
 &\longrightarrow_{\beta\eta}\mathbf{T}\underline{n}(\lambda z. 
   FG(A(\underline{n}\,*\,\<z\>))(B(\underline{n}\,*\,\<z\>))z)
  \underline{m}\cr
 &\longrightarrow_{\beta\eta}^{*}
  (\lambda z. FG(A(\underline{n}\,*\,\<z\>))(B(\underline{n}\,*\,
  \<z\>))z) \underline{m}\cr
 &\longrightarrow_{\beta\eta}
  FG(A(\underline{n}\,*\,\<\underline{m}\>))(B(\underline{n}\,*\,
  \<\underline{m}\>))\underline{m}\cr
  }
\]
and reducing in a similar way $B \underline{n} \underline{m}$, we
see that
\[FG(A(\underline{n}\,*\,\<\underline{m}\>))(B(\underline{n}\,*\,
\<\underline{m}\>))\underline{m}
 =_{\omega} FG(B(\underline{n}\,*\,
  \<\underline{m}\>))(A(\underline{n}\,*\,\<\underline{m}\>))\underline{m}
\]
has a proof of the same (smaller) ordinal. Thus, by Corollary \ref{41},
\[A(\underline{n}\,*\,\<\underline{m}\>) =_{\omega}
 B(\underline{n}\,*\,\<\underline{m}\>)
\]
  has a proof with the same or smaller ordinal.

  Thus by induction hypothesis, the extension of $n* \ \<m\>$ in the
  tree is well-founded.  So, every extension of $n$ in the tree is
  well-founded. Thus the subtree rooted at $n$ is well-founded. This
  contradicts the choice of $n$.\smallskip

{\em Case 3}. $\mathcal{T}$ has an endpiece.
Now, by Proposition \ref{application}, for each $m$ there exist term
$R_1, \ldots, R_k$ such that we have a
canonical proof, with an endpiece of the same rank as $\mathcal{T}$, of
\[\eqalign{A\underline{n}\underline{m} R_1 \cdots R_k
 &\longrightarrow_{\beta\eta}^{*} (\lambda z. FG(A(\underline{n}\,*\,
  \<z\>))(B(\underline{n}\,*\,\<z\>))z)\underline{m}R_1 \cdots R_k\cr
 &\longrightarrow_{\beta\eta}^{*}\dots\;
  _{\beta\eta}^{*}\!\longleftarrow B \underline{n} \underline{m}R_1
  \cdots R_k\ .\cr
  }
\]
Now consider that this endpiece is $\mathcal{X}$-canonical.\\
So, to equalize 
\[FG(A(\underline{n}\,*\,\<\underline{m}\>))(B(\underline{n}\,*\,\<\underline{m}\>))\underline{m}R_1
\cdots R_k
\]
 with
\[FG(B(\underline{n}\,*\,\<\underline{m}\>))(A(\underline{n}\
 * \ \<\underline{m}\>))\underline{m}R_1 \cdots R_k\ ,
\]
 it is necessary that some of instances of the $\omega$-rule,
 occurring in the endpiece, supplies a proof of
 $A(\underline{n}\,*\,\<\underline{m}\>) =_{\omega}
 B(\underline{n}\,*\,\<\underline{m}\>)$.

To see this consider the particular case when there is only one leaf
which is a direct conclusion of the $\omega$-rule.
\[\eqalign{A\underline{n}\underline{m}R_1 \cdots R_k
 &\longrightarrow_{\beta\eta}^{*} (\lambda z. FG(A(\underline{n}\,*\,
  \<z\>))(B(\underline{n}\,*\,\<z\>))z)\underline{m}R_1 
  \cdots R_k \longrightarrow_{\beta\eta}^{*}\cr
 &\longrightarrow_{\beta\eta}^{*} H \;
  _{\beta\eta}^{*}\!\longleftarrow LQ =_{\omega}LR
  \longrightarrow_{\beta\eta}^{*}
  H^{+}\; _{\beta\eta}^{*}\!\longleftarrow\cr
 &_{\beta\eta}^{*}\!\longleftarrow \ (\lambda
  z.FG(B(\underline{n}\,*\,\<z\>))(A(\underline{n}\,*\,\<z\>))z)
  \underline{m}R_1 \cdots R_k \;_{\beta\eta}^{*}\!\longleftarrow\cr
 &_{\beta\eta}^{*}\!\longleftarrow B \underline{n}
\underline{m}R_1 \cdots R_k\ .
  }
\]
Since the endpiece is $\mathcal{X}$-canonical, it follows that $LQ$
has the form of $FG(A(\underline{n}\,*\,\<m\>))(B(\underline{n}\,*\,
\<m\>))\underline{m}R_1 \cdots R_k$ and $LR$ has the form of
$FG(B(\underline{n}\,*\,\<m\>))(A(\underline{n}\,*\,\<m\>))\underline{m}R_1
\cdots R_k$.

Now let $N$ be any term. By the definition of canonical proof, there
is a $\mathcal{X}$-canonical proof, with ordinal less than
$\mathcal{T}$, of $LQN =_{\omega} LRN$ and therefore a proof with
ordinal less than $\mathcal{T}$, of
\[FG(A(\underline{n}\,*\,\<m\>))(B(\underline{n}\,*\,
\<m\>))\underline{m}R_1 \cdots R_k N =_{\omega}
FG(B(\underline{n}\,*\,\<m\>))(A(\underline{n}\,*\,\<m\>))\underline{m}R_1
\cdots R_k N\ .
\]
 Again by Lemma \ref{cof2}, Lemma \ref{cof1} and Proposition \ref{plot4},
$A(\underline{n}\,*\,<\underline{m}>) =_{\omega}
 B(\underline{n}\,*\,<\underline{m}>)$
has a proof with the same or smaller ordinal as
\[FG(A(\underline{n}\,*\,\<m\>))(B(\underline{n}\,*\,
\<m\>))\underline{m}R_1 \cdots R_k N =_{\omega}
FG(B(\underline{n}\,*\,\<m\>))(A(\underline{n}\,*\,\<m\>))\underline{m}R_1
\cdots R_k N\ .
\]
Thus, by induction hypothesis, the extension of $n* \ \<m\>$ in the
tree is well-founded. Thus every extension of $n$ in the tree is
well-founded and again we contradict
 the choice of $n$.

The case with multiple leaves can be treated by induction on the
number of leaves, in the endpiece, that are direct conclusions of the
$\omega$-rule.

Considering such leaves from left to right, and using the fact that
the endpiece is $\mathcal{X}$-canonical
\begin{enumerate}[$\bullet$]
\item when the left hand side and the right hand side of the leaf
have both the form:
\[FG(A(\underline{n}\,*\,\<\underline{m}\>))(B(\underline{n}\,
  *\, \<\underline{m}\>))\underline{m}R_1 \cdots R_k
\]
then we move to the next leaf;
 \item at some leaf, we must have that
the left hand side has the form
\[FG(A(\underline{n}\,*\,\<\underline{m}\>))(B(\underline{n}\,
  *\,\<\underline{m}\>))\underline{m}R_1 \cdots R_k
\]
 and the right hand side of the leaf has the form
\[FG(B(\underline{n}\,*\,\<m\>))(A(\underline{n}\,*\,\<m\>))
  \underline{m}R_1 \cdots R_k\ ,
\]
this case is treated as the one above.
\end{enumerate}
 This completes the proof.\qed

We have thus proved:
\begin{prop}\label{final} $A\underline{n} =_{\omega} B\underline{n}$ \iff\
the subtree $\mathbf{t}(n)$ rooted at $n$ is well-founded or $n$ is
not in the tree $\mathbf{t}$.\qed
\end{prop}

\begin{prop}{\bf (Main Theorem)} The set $\{ (M, N) |  M =_{\omega} N \}$ is
$\mathbf{\Pi_{1}^{1}}$-complete. \end{prop} \proof It easy to see
that equality in $\lambda\omega$ is $\mathbf{\Pi_{1}^{1}}$. On the
other hand, given any recursive tree $\mathbf{t}$ construct the
terms $A$ and $B$ (observe that the construction is effective and
uniform on (the term $\mathbf{T}$ representing) $\mathbf{t}$). Then
use Proposition \ref{final} to determine ({\em via}
equality in $\lambda\omega$) if $\mathbf{t} = \mathbf{t}(0)$ is
well-founded.
\qed

\section*{Acknowledgements}
We thank all the anonymous referees for giving substantial help in
improving a previous version of the paper.
 

\begin{thebibliography}{10}
 \bibitem{Ba84}    H.P. Barendregt. {\em The Lambda Calculus. Its Syntax and
 Semantics.} North-Holland, 1984.
 \bibitem{Boh} C. B\"{o}hm (Editor). {\em  $\lambda$-Calculus
 and Computer Science Theory}. LNCS 37, Springer 1975.
\bibitem{Cantini} A. Cantini.  Remarks on Applicative Theories. {\em Annals of Pure and
 Applied Logic} {\bf 136} (2005) pp. 91-115.
 \bibitem{FlaggMyhill} R.C. Flagg, J. Myhill.  Implication and Analysis
 in Classical Frege Structure. {\em Annals of Pure and
 Applied Logic} {\bf 34} (1987) pp.33-85.
 \bibitem{Rogers} H.jr Rogers. {\em Theory of Recursive Functions and
 Effective Computability}. MacGraw Hill New  York 1967.
\bibitem{IS2004} B. Intrigila, R. Statman.
The Omega Rule is $\mathbf{\Pi}^{0}_{2}$-Hard in the $\lambda\beta
$-Calculus. {\em LICS 2004} pp.202-210, IEEE Computer Society 2004.
\bibitem{IS2002} B. Intrigila, R. Statman. Some Results on Extensionality in Lambda
Calculus. {\em Annals of Pure and Applied Logic}. {\bf 132}, Issues
2-3, (2005) pp.109-125.
\bibitem{IS2006} B. Intrigila, R. Statman. Solution of a Problem of
Barendregt on Sensible $\lambda$-Theories. {\em Logical Methods in
Computer Science} {\bf 2} (2006), Issue 4.
\bibitem{IS2007} B. Intrigila, R. Statman.
The Omega Rule is $\mathbf{\Pi}^{1}_{1}$-Complete in the
$\lambda\beta $-Calculus. {\em TLCA 2007} pp.178-193, LNCS 2007.
\bibitem{Plo} G. Plotkin. The $\lambda$-Calculus is $\omega$-incomplete.
{\em J. Symbolic Logic}, 39, pp. 313-317.
\bibitem{Sch} K. Sch\"{u}tte. {\em Proof Theory}.
                    Springer Verlag New  York Heidelberg Berlin 1977.
\bibitem{St1} R. Statman. Gentzen's Notion of a Direct Proof.
                    In {\em Handbook of Mathematical Logic}. (K.J. Barwise Editor)
                    North Holland Amsterdam 1978.
 \bibitem{St2} R. Statman. Normal Varieties of Combinators. In
  {\em  Logic from Computer Science}. (Y.N. Moschovakis Editor) Springer 1992.

 \end{thebibliography}
\end{document}